\documentclass[12pt]{article}
\pdfoutput=1
\usepackage{hyperref}
\usepackage{amsmath,amssymb,amsthm,amscd}
\usepackage{mathtools}
\usepackage{bm}
\usepackage{graphicx}
\hypersetup{pdftitle={Superconformal Index Identities for Dual Gauge Theories},
pdfauthor={Min-xin Huang, Gao-fu Ren, Xin Wang}}

\newcommand{\dd}{\mathrm{d}}

\newcommand{\Tr}{\textrm{Tr}}

\newcommand{\be}{\begin{equation}}
\newcommand{\ee}{\end{equation}}
\newcommand{\ba}{\begin{aligned}}
\newcommand{\ea}{\end{aligned}}

\numberwithin{equation}{section}

\newtheorem{theorem}{Theorem}[section]

\newtheorem{lemma}[theorem]{Lemma}
\newtheorem{corollary}[theorem]{Corollary}
\theoremstyle{definition}

\theoremstyle{remark}

\begin{document}
\hypersetup{pageanchor=false}
\begin{titlepage}
\noindent\hfill USTC-ICTS/PCFT-26-94\par
\vskip 3 cm
\begin{center}
{\Large\bfseries Superconformal Index Identities\\[4pt]
for Dual Gauge Theories\par}
\end{center}
\vskip 0.5 cm
\renewcommand{\thefootnote}{\fnsymbol{footnote}}
\vskip 30pt
{\centering\large
Min-xin Huang\footnote{\href{mailto:minxin@ustc.edu.cn}{\nolinkurl{minxin@ustc.edu.cn}}},
Gao-fu Ren\footnote{\href{mailto:rengaofu@mail.ustc.edu.cn}{\nolinkurl{rengaofu@mail.ustc.edu.cn}}}, and
Xin Wang\footnote{\href{mailto:wxin@ustc.edu.cn}{\nolinkurl{wxin@ustc.edu.cn}}}\par}
\vskip .5cm \vskip 20pt
\begin{center}
{Interdisciplinary Center for Theoretical Study,\\ \vskip 0.1cm University of Science and Technology of China, Hefei, Anhui 230026, China}
\\ \vskip 0.3 cm
{Peng Huanwu Center for Fundamental Theory,\\ \vskip 0.1cm Hefei, Anhui 230026, China}
\end{center}
\setcounter{footnote}{0}
\renewcommand{\thefootnote}{\arabic{footnote}}
\vskip 40pt
\begin{abstract}
We study superconformal index identities for three classes of dual
four-dimensional gauge theories. For $\mathcal N=4$ super-Yang--Mills
theory with gauge groups $SO(2n+1)$ and $USp(2n)$, we prove at every
rank the equality of flavored Schur indices predicted by $S$-duality,
using Rains's diagonal trace transformation. For Kutasov--Schwimmer
duality with adjoint superpotential $\Tr X^{k+1}$, we prove the index
identity between $SU(k+1)$ and $SU(k-1)$ with two flavors at
independent elliptic nomes $p,q$, in the ultraviolet assignment
$R(Q)=0$. On the locus $p=q^k$, we reduce the identity to that of
ordinary supersymmetric QCD whenever the specialized expressions are
generically regular. Finally, we test the dualities conjectured by
Cho and Song between quiver and single-group descriptions of four
families of Argyres--Douglas theories. For the rank-one $H_1$ theory
we prove equality of the two full flavored indices, and for each
family at arbitrary rank we evaluate the generalized Coulomb index in
both descriptions, obtaining the product formula conjectured by
Buican and Jiang.
\end{abstract}
\end{titlepage}
\hypersetup{pageanchor=true}
\vfill \eject
\newpage
\baselineskip=16pt
\tableofcontents

\section{Introduction}\label{sec:introduction}

Seiberg's discovery of electric--magnetic duality in supersymmetric
QCD was a landmark in the study of strongly coupled gauge theories
\cite{Seiberg:1994pq}. In the conformal window $3N/2<F<3N$,
four-dimensional $\mathcal N=1$ supersymmetric QCD with gauge group
$SU(N)$ and $F$ flavor pairs has a magnetic description with gauge
group $SU(F-N)$. An electric meson, built from a quark and an
antiquark, becomes an elementary gauge singlet in the magnetic theory.
Seiberg's identification of operators across this change of variables
turned global symmetries, deformations and renormalization-group flows
into quantitative tests of a proposed infrared equivalence.

The superconformal index extends these tests to an infinite set of
protected states. Building on Witten's observation that supersymmetry
pairs bosonic and fermionic excitations \cite{Witten:1982df},
R\"omelsberger and, independently, Kinney, Maldacena, Minwalla and
Raju constructed indices for four-dimensional superconformal theories
\cite{Romelsberger:2005eg,Kinney:2005ej}. In radial quantization the states on the
three-sphere correspond to local operators. Fugacities keep track of
their commuting charges, and cancellations between paired states make
the signed count insensitive to continuous changes of the couplings
that preserve supersymmetry. A calculation at weak coupling can
therefore test a proposed equivalence at strong coupling: dual
descriptions must have equal indices, although long multiplets are
invisible to this comparison. For reviews, see
\cite{Rastelli:2014jja,Rastelli:2016tbz,Gadde:2020yah}.

For gauge theories the index is an elliptic hypergeometric integral,
so a duality predicts a transformation between two such integrals.
Spiridonov's elliptic beta integral gives the identity for the
confining $SU(2)$ example \cite{MR1846786}. Rains subsequently
proved a transformation of multivariate elliptic hypergeometric
integrals \cite{Rains:2005lkc}, and Dolan and Osborn recognized its
two sides as the electric and magnetic indices of Seiberg-dual SQCD
\cite{Dolan:2008qi}. An elegant feature of their identification is
that one formula produces both the change in gauge rank and the
elliptic gamma factors of the magnetic mesons. Dolan and Osborn also
considered symplectic and orthogonal gauge groups, and Spiridonov and
Vartanov brought a much wider collection of dualities into this
correspondence \cite{Spiridonov:2009za}. Their work makes it possible to read a
proposed spectrum from the factors of an integral identity and,
conversely, to use a proposed duality to predict a transformation.
More recently, Kels and Yamazaki derived lens-space identities for
orthogonal Seiberg-dual pairs \cite{Kels:2026axw}.

We follow this approach for three four-dimensional problems: the
flavored Schur indices of $\mathcal N=4$ theories with gauge algebras
$B_n$ and $C_n$, finite-rank Kutasov--Schwimmer duals with one
adjoint, and quiver and single-group descriptions of Argyres--Douglas
theories. Section~\ref{sec:tools} collects the notation and the
integral transformations used below.

Our first example is $\mathcal N=4$ super-Yang--Mills theory with
gauge algebras $B_n$ and $C_n$. Goddard, Nuyts and Olive identified
the dual group from the allowed magnetic charges
\cite{Goddard:1976qe}. Combined with Montonen and Olive's proposal
to exchange electric and magnetic charges \cite{Montonen:1977sn}, this
predicts a duality between $SO(2n+1)$ and $USp(2n)$. Kapustin and
Witten showed how the exchange of Wilson and 't~Hooft operators in a
topological twist gives a gauge-theoretic explanation of geometric
Langlands duality \cite{Kapustin:2006pk}. Gaiotto and Witten's
pioneering analysis of half-BPS boundary conditions revealed a link
between four-dimensional $S$-duality and 3d mirror symmetry. By
following the branes through the duality, they identified the degrees
of freedom at the dual boundary \cite{Gaiotto:2008ak}.

Gadde, Pomoni, Rastelli and Razamat wrote down the $B_n$ and $C_n$
index integrals and compared their series expansions
\cite[Appendix~A]{Gadde:2009kb}, and Spiridonov and
Vartanov proved the equality of the two elliptic indices in several
important special cases \cite{Spiridonov:2010qv}. The work of Gadde, Rastelli,
Razamat and Yan on limits of the $\mathcal N=2$ index revealed the
role of symmetric polynomials in the Schur index of local operators
\cite{Gadde:2011uv}, and subsequent work treated the Schur and deformed Schur
indices for the classical gauge groups \cite{Du:2023kfu,Ren:2025tvx}. For the
unflavored Schur index, Kulanthaivelu checked and presented an
all-rank operator proof based on an argument generated by GPT-5.6 Sol
\cite{Kulanthaivelu}. Li and Pan obtained equal unflavored expressions
at rank three and checked the flavored equality by series expansion
\cite[Sec.~4.3]{Li:2026bch}. In Section~\ref{sec:bc-schur} we prove the
flavored equality at every rank by identifying the two Schur integrals
with the two sides of a specialization of Rains's diagonal trace
transformation \cite{MR3772016}. The proof retains the full flavor
fugacity and includes the normalization and the boundary limit needed
to recover the physical integration contour. The unflavored equality
is a direct corollary.

Kutasov's cubic-adjoint duality \cite{Kutasov:1995ve} led to an
elegant generalization by Kutasov and Schwimmer \cite{Kutasov:1995np}.
For $SU(N)$ with $F$ flavor pairs, one adjoint $X$ and
$W=\operatorname{Tr}X^{k+1}$, they recognized that the same integer
$k$ fixes the magnetic color number $M=kF-N$ and counts the electric
meson multiplets $\widetilde Q X^j Q$, $0\leq j<k$, that become
elementary singlets in the magnetic theory. Intriligator, Leigh and
Strassler traced related dualities through symmetry breaking and
confinement in a wider class of chiral and nonchiral theories
\cite{Intriligator:1995ax}. Returning to the adjoint theory,
Kutasov, Schwimmer and Seiberg followed the mesons under polynomial
deformations and showed how quantum relations reconcile the electric
and magnetic chiral rings \cite{Kutasov:1995ss}. On the index
side, Dolan and Osborn verified the adjoint index matching in the
Veneziano limit $N,F\to\infty$ with $F/N$ fixed
\cite[Sec.~4]{Dolan:2008qi}, and Kutasov and Lin turned this check into
a way to recover the multiplicities and $R$-charges of the magnetic
mesons from the large-$N$ index \cite[Sec.~2]{Kutasov:2014wwa}.

In Section~\ref{sec:ks} we derive two restricted forms of the
finite-rank identity from established transformations. The first
holds at independent nomes $p,q$ for $F=2$, $N=k+1$ and $M=k-1$, in an
ultraviolet grading with $R(Q)=0$; relating it to the infrared
superconformal index would require a separate treatment of free
operators and accidental symmetries. The second reduces the identity
on the locus $p=q^k$ to ordinary SQCD, for balanced flavor families on
which the specialized expressions are generically regular. Both are
exact meromorphic identities within their stated domains.

The discovery by Argyres and Douglas that mutually nonlocal charged
particles can become massless at the same vacuum revealed a new class
of interacting four-dimensional $\mathcal N=2$ fixed points
\cite{Argyres:1995jj}. Xie put a large family of these theories on a
geometric footing, constructing them from six dimensions with
irregular punctures and reading off their Coulomb branch spectra from
the Seiberg--Witten curves \cite{Xie:2012hs}. This geometry made new
index calculations possible. Buican and Nishinaka used two-dimensional
$q$-deformed Yang--Mills theory to propose closed expressions for the
Schur indices of the $(A_1,A_{2n-3})$ and $(A_1,D_{2n})$ series
\cite{Buican:2015ina}. Maruyoshi and Song used $\mathcal N=1$
ultraviolet Lagrangians to compute the full indices of the $(A_1,A_m)$
theories \cite{Maruyoshi:2016aim}, and together with Agarwal they extended
this approach to the $(A_1,D_m)$ series \cite{Agarwal:2016pjo}.

For a class of theories engineered from M5-branes, Song, Xie and Yan
matched Schur indices with vacuum characters of the proposed
two-dimensional chiral algebras \cite{Song:2017oew}. Xie and Yan
showed how the resulting $W$-algebra character formulas expose
level-rank dualities directly in the Schur spectrum \cite{Xie:2019zlb}.
In the generalized Coulomb limit, Buican and Jiang conjectured a
product formula determined by the scaling dimensions of the generators
of a freely generated Coulomb branch chiral ring
\cite[Conjecture~4]{Buican:2023efi}.

Cho and Song proposed an ingenious quiver construction for four
Argyres--Douglas families \cite{Cho:2026tqf}, which encodes the rank $n$
of the infrared theory in a chain of $n$ gauge nodes, each with gauge
group $SU(2)$. For $n>1$, their conjecture predicts an identity
between a single higher-rank integral and a chain of rank-one
integrals. In Section~\ref{sec:ad-quivers} we prove the full flavored
identity for the rank-one $H_1$ theory by specializing van de Bult's
transformation \cite{MR2787288}. At arbitrary rank we evaluate both
descriptions of all four families in the generalized Coulomb limit
$q\to0$ with $p$ and $\xi=pq/t$ fixed. The Macdonald constant-term
theorem \cite{MR1224114,MR2199453} gives the
product predicted by Buican and Jiang, and setting $p=0$ recovers the
Coulomb branch generating functions for the spectra identified in
\cite{Cho:2026tqf}.

Section~\ref{sec:conclusion} discusses the open questions at generic
fugacities, further four-dimensional examples, and extensions to
defects, boundaries and 3d mirror symmetry.

\section{Preliminaries}
\label{sec:tools}

\subsection{The index and three kinds of identities}

We use the four-dimensional $\mathcal N=1$ index
\cite{Romelsberger:2005eg,Kinney:2005ej,Dolan:2008qi} in the normalization
\begin{equation}
 \mathcal I(p,q;\boldsymbol x)
 =\operatorname{Tr}_{\delta=0}(-1)^{\mathsf F}
 p^{j_1+j_2+R/2}q^{-j_1+j_2+R/2}
 \prod_a x_a^{f_a},\qquad
 \delta=E-2j_2-\frac32R.
 \label{eq:tools-index}
\end{equation}
Here $E$ is the energy, $j_1,j_2$ are the spins, $R$ is the $R$-charge,
$f_a$ are flavor charges, and $\mathsf F$ is the fermion number; the
italic $F$ will denote a number of flavors. The index receives
contributions only from states with $\delta=0$.

We distinguish three kinds of statements. A \emph{full-index identity}
retains all the fugacities specified in its theorem; an identity on a
\emph{special locus}, such as $p=q^k$, holds only on that locus; and a
\emph{limit identity}, such as a Coulomb-limit identity, compares only
the terms that survive the limit. In particular, agreement at $p=q^k$
for fixed $k$ does not imply agreement at independent $p,q$. Each
theorem below is a statement about the integrals, independent of
whether the corresponding duality holds.

\subsection{Common notation}

For the four-dimensional index integrals we write
\begin{equation}
 |p|,|q|<1,\qquad P=pq,\qquad
 \kappa(p,q)=(p;p)_\infty(q;q)_\infty,
 \label{eq:tools-nomes}
\end{equation}
so $p,q$ are the elliptic nomes and $P$ always denotes their product.
Frequently used symbols are listed below.
\begin{center}
\begin{tabular}{p{0.20\textwidth}p{0.70\textwidth}}
\hline
Symbol & Meaning \\
\hline
$N,M$ & Electric and magnetic color numbers in unitary theories \\
$F$ & Number of fundamental--antifundamental pairs in unitary theories \\
$k$ & Integer in the adjoint superpotential \\
$n$ & Rank for $B_n/C_n$; integer labeling the Argyres--Douglas families \\
$u,v$ & Multiplicative weights of adjoint chiral fields \\
$a_i,b_i$ & Fundamental and antifundamental weights,
including their R-charge powers \\
$A,B$ & Products $A=\prod_i a_i$, $B=\prod_i b_i$ \\
$z_i$ & Gauge variables to be integrated \\
$s,x$ & Schur expansion variable and $SU(2)$ flavor fugacity in
Section~\ref{sec:bc-schur}, with $u=sx$, $v=s/x$ \\
$t,\xi$ & $\mathcal N=2$ fugacity and ratio $\xi=P/t$ retained in the AD limits \\
$\tau$ & Auxiliary Selberg parameter in integral transformations \\
$\lambda$ & A regulator used in a contour limit \\
$\rho$ & A new nome after a change of infinite-product bases \\
\hline
\end{tabular}
\end{center}
The symbol $t$ is reserved for the physical $\mathcal N=2$ fugacity and
$\tau$ for the auxiliary parameter of an integral transformation. The
Schur products of Section~\ref{sec:bc-schur} have nome $s^2$, and the
nomes of the elliptic transformation applied to them are written
$p_*,q_*$.

\subsection{Infinite products}

For $|q|<1$ we use
\begin{equation}
 (z;q)_\infty=\prod_{j=0}^{\infty}(1-zq^j),\qquad
 \theta(z;q)=(z;q)_\infty(q/z;q)_\infty,
 \label{eq:tools-pochhammer}
\end{equation}
and a list of arguments denotes a product, e.g.\
$(a,b;q)_\infty=(a;q)_\infty(b;q)_\infty$. The elliptic gamma
function is
\begin{equation}
 \Gamma(z;p,q)=\prod_{j,l=0}^{\infty}
 \frac{1-p^{j+1}q^{l+1}/z}{1-p^jq^lz}.
 \label{eq:tools-gamma}
\end{equation}
The nomes are suppressed when they do not change within a
calculation. We also write
\begin{equation}
 \Gamma(z_1,\ldots,z_m;p,q)=\prod_{a=1}^m\Gamma(z_a;p,q),
 \qquad
 G(c;z)=\Gamma(cz;p,q)\Gamma(c/z;p,q).
 \label{eq:tools-G}
\end{equation}
Signs $\pm$ are independent, so $\Gamma(cz^{\pm1}w^{\pm1})$ is a
product of four factors. We use repeatedly the reflection and shift
identities
\begin{equation}
 \Gamma(z;p,q)\Gamma(P/z;p,q)=1,
 \label{eq:tools-reflection}
\end{equation}
\begin{equation}
 \Gamma(pz;p,q)=\theta(z;q)\Gamma(z;p,q),\qquad
 \Gamma(qz;p,q)=\theta(z;p)\Gamma(z;p,q),
 \label{eq:tools-shift}
\end{equation}
and the residue at $z=1$,
\begin{equation}
 \lim_{z\to1}(1-z)\Gamma(z;p,q)=\kappa(p,q)^{-1}.
 \label{eq:tools-residue}
\end{equation}

\subsection{Chiral multiplets and gauge integrals}

The plethystic exponential is
\begin{equation}
 \operatorname{PE}[f(p,q,z)]
 =\exp\left(\sum_{m=1}^{\infty}\frac1m
 f(p^m,q^m,z^m)\right).
 \label{eq:tools-PE}
\end{equation}
The single-letter index of a chiral multiplet of weight $z$ is
\begin{equation}
 f_{\mathrm{chiral}}(p,q,z)
 =\frac{z-P/z}{(1-p)(1-q)},
 \label{eq:tools-letter}
\end{equation}
and
\begin{equation}
 \operatorname{PE}\left[\frac{z-P/z}{(1-p)(1-q)}\right]
 =\prod_{j,l\geq0}
 \frac{1-p^{j+1}q^{l+1}/z}{1-p^jq^lz}
 =\Gamma(z;p,q)
 \label{eq:tools-chiral}
\end{equation}
for $|P|<|z|<1$, and elsewhere by continuation. A chiral field of
$R$-charge $R_0$ and flavor weight $x$ has $z=P^{R_0/2}x$; we include
the $R$-charge factors in the parameters $a_i,b_i,u,v$.

A field in a representation contributes one gamma factor for each
weight. For $SU(N)$, with $\prod_jz_j=1$, the fundamental has weights
$z_j$, the antifundamental $z_j^{-1}$, and the adjoint the nonzero
weights $z_i/z_j$, $i\ne j$, together with $N-1$ zero weights, so that
an adjoint of weight $u$ contributes
\begin{equation}
 \Gamma(u)^{N-1}\prod_{i\ne j}\Gamma(uz_i/z_j).
 \label{eq:tools-adjoint}
\end{equation}
Including the vector multiplet, the normalized $SU(N)$ measure is
\begin{equation}
 \dd\mu_N(\boldsymbol z;p,q)
 =\frac{\kappa(p,q)^{N-1}}{N!}
 \frac{\displaystyle\prod_{j=1}^{N-1}
       \dd z_j/(2\pi\mathrm i z_j)}
      {\displaystyle\prod_{i\ne j}\Gamma(z_i/z_j;p,q)},
 \qquad z_N=(z_1\cdots z_{N-1})^{-1}.
 \label{eq:tools-measure}
\end{equation}
For $N=1$ the integral is evaluation at $z_1=1$. The basic integrals,
for SQCD and for SQCD with one adjoint, are
\begin{align}
 \mathcal S_{N,F}(\boldsymbol a,\boldsymbol b;p,q)
 &=\int\dd\mu_N(\boldsymbol z;p,q)
   \prod_{i=1}^F\prod_{j=1}^N
       \Gamma(a_i z_j,b_i/z_j;p,q),
 \label{eq:tools-S}\\
 \mathcal A_{N,F}(u;\boldsymbol a,\boldsymbol b;p,q)
 &=\Gamma(u;p,q)^{N-1}\int\dd\mu_N(\boldsymbol z;p,q)
   \prod_{i\ne j}\Gamma(uz_i/z_j;p,q)\notag\\
 &\hspace{5mm}\times\prod_{i=1}^F\prod_{j=1}^N
       \Gamma(a_i z_j,b_i/z_j;p,q).
 \label{eq:tools-A}
\end{align}
The balancing conditions imposed on their parameters differ between
theorems.

For a root $\alpha=(\alpha_1,\ldots,\alpha_n)$ of a root system $\Phi$
we write $\boldsymbol z^\alpha=\prod_i z_i^{\alpha_i}$, and
$\operatorname{CT}_{\boldsymbol z}$ denotes the coefficient of
$z_1^0\cdots z_n^0$ in a convergent Laurent expansion, equivalently
the integral over $|z_1|=\cdots=|z_n|=1$ with measure
$\prod_i\dd z_i/(2\pi\mathrm i z_i)$.

\subsection{Contours and analytic continuation}

The gamma function \eqref{eq:tools-gamma} has poles at
$z=p^{-j}q^{-l}$ and zeros at $z=p^{j+1}q^{l+1}$, $j,l\geq0$. The
poles of $\Gamma(a/z)$ accumulate at $0$ and those of
$\Gamma(bz)$ at infinity, and an admissible contour separates the two
families; for $|a|,|b|<1$ the unit circle suffices.

On a parameter domain where a fixed compact contour is admissible,
locally uniform convergence makes the integral holomorphic in the
parameters. Beyond it, the integrals are continued meromorphically
with the contours of the quoted transformations, and pole crossings
are accounted for by residues \cite[Appendix~10]{Rains:2005lkc}:
\[
 I_{\mathrm{cont}}(\boldsymbol t)
 =I_{C'}(\boldsymbol t)+\sum_\nu R_\nu(\boldsymbol t),
\]
where $C'$ avoids the moving poles and the residue contributions
$R_\nu$ include, for multiple integrals, lower-dimensional integrals,
whose continuation must be justified by the relevant transformation
or contour argument, not by meromorphy of the integrand alone.
Fractional powers are
taken on compatible branches, or on the corresponding parameter
cover. An equality on a nonempty open subset of the balancing variety
then holds on the common connected domain of the continuations;
equality on a proper subvariety, such as $p=q^k$, does not extend in
this way.

A contour is \emph{pinched} when poles from the two families approach
the same point. In the applications below a prefactor such as
$1/\Gamma(\lambda^{-1})$ tends to zero while the integral develops a
compensating pole, and the limit $\lambda\to1$ is taken of their
product. Interchanges of integration are justified by Fubini's
theorem on a domain of absolute integrability, and limits of integrals
by dominated convergence, before any analytic continuation.

\subsection{Rains's \texorpdfstring{$A_n$}{An} transformation}

Rains's $A_n$ transformation \cite[Theorem~4.1]{Rains:2005lkc}, which
contains Spiridonov's elliptic beta integral \cite{MR1846786} as
a special case, is the index identity of Seiberg duality
\cite{Dolan:2008qi,Spiridonov:2009za}. In the normalization \eqref{eq:tools-S} it
reads as follows.

\begin{theorem}[Ordinary unitary transformation]
\label{thm:tools-SQCD}
Let $N,F$ be positive integers, $M=F-N\geq1$, and
\begin{equation}
 A=\prod_{i=1}^F a_i,\qquad B=\prod_{i=1}^F b_i,
 \qquad AB=P^M.
 \label{eq:tools-balance}
\end{equation}
Choose the roots so that $A^{1/M}B^{1/M}=P$.
The following meromorphic identity holds:
\begin{align}
 \mathcal S_{N,F}(\boldsymbol a,\boldsymbol b;p,q)
 ={}&\prod_{i,j=1}^F\Gamma(a_i b_j;p,q)
 \mathcal S_{M,F}\left(
    \left\{\frac{A^{1/M}}{a_i}\right\}_{i=1}^F,
    \left\{\frac{B^{1/M}}{b_i}\right\}_{i=1}^F;p,q\right).
 \label{eq:tools-SQCD}
\end{align}
\end{theorem}

Only the product of the two roots is fixed. Another choice multiplies
$A^{1/M}$ by an $M$th root of unity and $B^{1/M}$ by its inverse,
which is absorbed by rescaling the magnetic gauge variables. For
$F=N+1$ the magnetic integral is an evaluation.
\begin{corollary}[Confining evaluation]
\label{cor:tools-confining}
If $F=N+1$ and $AB=P$, then
\begin{equation}
 \mathcal S_{N,N+1}(\boldsymbol a,\boldsymbol b;p,q)
 =\prod_{i,j=1}^{N+1}\Gamma(a_i b_j;p,q)
  \prod_{i=1}^{N+1}\Gamma(A/a_i,B/b_i;p,q).
 \label{eq:tools-confining}
\end{equation}
\end{corollary}
For $N=2$ this is Spiridonov's six-parameter evaluation.

\subsection{Rains's diagonal trace transformation}
\label{subsec:tools-rains-trace}

Van de Bult's univariate transformation \cite[Theorem~3.2]{MR2787288}
was extended to arbitrary rank by Rains
\cite[Theorem~3.22]{MR3772016}. We state Rains's theorem and then
recover van de Bult's result, in its original normalization, as the
case $n=1$.

In this subsection we keep Rains's notation: $t$ is the Selberg
parameter, unrelated to the $\mathcal N=2$ fugacity, and
$u_0,\ldots,u_3$ are boundary variables; in
Section~\ref{sec:bc-schur} they become $\tau$ and
$\eta_0,\ldots,\eta_3$. For $0<|p|,|q|<1$ the elliptic Selberg
density, including the measure, is
\begin{equation}
\begin{split}
 \Delta_S^{(n)}(\boldsymbol z;a_0,\ldots,a_7;t;p,q)
 ={}&\frac{\kappa(p,q)^n\Gamma(t;p,q)^n}{2^n n!}
 \prod_{1\leq i<j\leq n}
 \frac{\Gamma(tz_i^{\pm1}z_j^{\pm1};p,q)}
      {\Gamma(z_i^{\pm1}z_j^{\pm1};p,q)}\\
 &\times\prod_{i=1}^n
 \frac{\prod_{r=0}^7\Gamma(a_rz_i^{\pm1};p,q)}
      {\Gamma(z_i^{\pm2};p,q)}
 \prod_{i=1}^n\frac{\dd z_i}{2\pi\mathrm i z_i}.
\end{split}
\label{eq:tools-rains-selberg-density}
\end{equation}
Let $\mathcal K_c^{(n)}(\boldsymbol z;\boldsymbol w;t;p,q)$
be Rains's analytic interpolation kernel
\cite[Definition~2.2 and Theorem~3.1]{MR3772016}.
The special value needed below is
\cite[Proposition~2.10]{MR3772016}
\begin{equation}
 \mathcal K_{\sqrt{pq/t}}^{(n)}
 (\boldsymbol z;\boldsymbol w;t;p,q)
 =\prod_{i,j=1}^n
 \Gamma(\sqrt{pq/t}\,z_i^{\pm1}w_j^{\pm1};p,q).
 \label{eq:tools-rains-kernel-product}
\end{equation}

\begin{theorem}[Rains's diagonal trace transformation
 {\cite[Theorem~3.22]{MR3772016}}]
\label{thm:tools-rains-trace}
For a positive integer $n$ and generic nonzero $t,c,u_0,u_1,u_2,u_3$,
the integral
\begin{multline}
 \int\mathcal K_c^{(n)}(\boldsymbol z;\boldsymbol z;t;p,q)\\
 \times\Delta_S^{(n)}\!\left(
 \boldsymbol z;
 \sqrt{pq/c}\,u_0^{\pm1},
 \sqrt{pq/c}\,u_1^{\pm1},
 \sqrt{pq/c}\,u_2^{\pm1},
 \sqrt{pq/c}\,u_3^{\pm1};t;p,q\right)
 \label{eq:tools-rains-trace-integral}
\end{multline}
is invariant under
\begin{equation}
 u_r\longmapsto
 \frac{u_r}{\sqrt{u_0u_1u_2u_3}},
 \qquad 0\leq r\leq3.
 \label{eq:tools-rains-trace-map}
\end{equation}
Square roots are chosen consistently on a local parameter domain.
The integrals use admissible separating contours and are then
continued meromorphically. No balancing condition on the $u_r$ is
required, and there is no scalar multiplier.
\end{theorem}

The specialization used in Section~\ref{sec:bc-schur} is
\begin{equation}
 (p,q,t,c)=(u,s^2,v,u),\qquad v=s^2/u,\qquad u_r=d_r/s,
 \label{eq:tools-rains-dictionary}
\end{equation}
where $u,v$ are the adjoint weights of that section. Then
$c^2=pq/t$, so the kernel is the product
\eqref{eq:tools-rains-kernel-product}, and the boundary
transformation becomes $d_r\mapsto s^2d_r/\sqrt{d_0d_1d_2d_3}$.

\paragraph{The rank-one case: van de Bult's transformation.}
The case $n=1$ of Theorem~\ref{thm:tools-rains-trace} is van de
Bult's transformation \cite[Theorem~3.2]{MR2787288}, as noted before
Theorem~3.22 of \cite{MR3772016}. We state it in van de Bult's
parameters and normalization.

With $\boldsymbol t=(t_1,t_2,t_3,t_4)$, van de Bult's normalization
\cite[Eq.~(5)]{MR2787288} uses the double products
\[
 (a_1,\ldots,a_m;p,q)_\infty
 =\prod_{\ell=1}^{m}\prod_{i,j\ge0}(1-a_\ell p^iq^j).
\]
Define
\begin{align}
 \mathcal N_b(\boldsymbol t;p,q)
 &=\frac{(P/b;p,q)_\infty^4
       (b^2,pb^2,qb^2,Pb^2;p,q)_\infty}
       {(b,pb,qb,Pb;p,q)_\infty}
 \notag\\
 &\quad\times\prod_{1\le r<s\le4}
 \left(t_rt_s,\frac{Pt_r}{bt_s},\frac{Pt_s}{bt_r},
       \frac{P^2}{b^2t_rt_s};p,q\right)_\infty,
 \label{eq:tools-bult-normalization}\\
 E(b;t_1,t_2,t_3,t_4;p,q)
 &=\mathcal N_b(\boldsymbol t;p,q)\,
   \frac{\kappa}{2}\oint_C\frac{\dd z}{2\pi\mathrm i z}
   \frac{\Gamma(bz^{\pm2})}{\Gamma(z^{\pm2})}
   \prod_{r=1}^{4}
   \frac{\Gamma(t_rz^{\pm1})}{\Gamma(bt_rz^{\pm1})}.
 \label{eq:tools-bult-E}
\end{align}
Initially take $0<|p|,|q|<1$ and generic nonzero $b,t_r$.
The positively oriented contour separates the inward pole sequences
of $\Gamma(bz^{\pm2})$,
$\Gamma(t_rz^{\pm1})$ and $\Gamma(Pz^{\pm1}/(bt_r))$
from their reciprocal outward sequences.
In particular, $C$ is the unit circle when
$|b|,|t_r|,|P/(bt_r)|<1$ for every $r$.
Other parameters are treated by analytic continuation.

\begin{corollary}[Van de Bult's rank-one transformation
 {\cite[Theorem~3.2]{MR2787288}}]
\label{thm:tools-bult}
With the preceding normalization,
\begin{equation}
 \begin{aligned}
 E(b;t_1,t_2,t_3,t_4;p,q)
 &=E(b;t_1v,t_2v,t_3v,t_4v;p,q),\\
 v^2&=\frac{p^2q^2}{b^2t_1t_2t_3t_4}.
 \end{aligned}
 \label{eq:tools-bult-transformation}
\end{equation}
Either sign of $v$ gives the same integral by $z\mapsto-z$.
\end{corollary}

\begin{proof}
For $n=1$, the diagonal interpolation kernel is
\cite[Corollary~2.8 and Theorem~3.1]{MR3772016}
\[
 \mathcal K_c^{(1)}(z;z;t;p,q)
 =\frac{\Gamma(cz^{\pm2};p,q)\Gamma(c;p,q)^2}
        {\Gamma(t;p,q)\Gamma(c^2;p,q)}.
\]
The factor $\Gamma(t;p,q)$ cancels against the rank-one Selberg
density, leaving the common scalar $\Gamma(c;p,q)^2/\Gamma(c^2;p,q)$,
which is independent of the boundary parameters. In particular, this
rank-one reduction does not require $c^2=pq/t$. Set
\[
 b=c,\qquad t_{r+1}=\sqrt{pq/c}\,u_r,\qquad 0\le r\le3.
\]
The reflection identity gives
$\Gamma(t_rz^{\pm1})/\Gamma(bt_rz^{\pm1})
=\Gamma(t_rz^{\pm1},pqz^{\pm1}/(bt_r))$,
so, after removing the common scalar, the trace integral is the
integral in \eqref{eq:tools-bult-E}. The boundary transformation
\eqref{eq:tools-rains-trace-map} becomes $t_r\mapsto vt_r$, with
\[
 v^2=\frac{1}{u_0u_1u_2u_3}
     =\frac{(pq)^2}{b^2t_1t_2t_3t_4}.
\]
Finally, $\mathcal N_b$ in \eqref{eq:tools-bult-normalization} is
invariant: its ratio terms are fixed, while the pair-product terms
are exchanged with the inverse-product terms for complementary
pairs. Multiplying by this invariant normalization gives
\eqref{eq:tools-bult-transformation}. The argument holds at generic
parameters and then extends meromorphically.
\end{proof}

In Section~\ref{sec:ad-quivers} we use it, in the form
\eqref{eq:ad-bult-map}, with
\begin{equation}
 b=c,\qquad t_r=h_r,\qquad v=\eta,
 \qquad
 \eta^2=\frac{P^2}{c^2h_1h_2h_3h_4}.
 \label{eq:tools-bult-dictionary}
\end{equation}

\section{The flavored \texorpdfstring{$B_n/C_n$}{Bn/Cn} Schur identity at arbitrary rank}
\label{sec:bc-schur}

In this section we prove that the flavored Schur indices of
$\mathcal N=4$ super-Yang--Mills theory with gauge groups $SO(2n+1)$
and $USp(2n)$ agree for every $n\geq1$, as predicted by $S$-duality.
The proof applies Rains's diagonal trace transformation
\cite[Theorem~3.22]{MR3772016} to an auxiliary elliptic gamma
integral whose first nome is an adjoint weight. With this choice the
shift and reflection identities reduce the integrand to theta
functions, and a theta duplication formula turns the boundary
parameters on the two sides into the roots that distinguish $B_n$
from $C_n$. It remains to match the
overall constants and to justify a limit of the contour. The argument
is uniform in the rank.

The full-index comparison was formulated in
\cite[Appendix~A]{Gadde:2009kb}. At ranks one and two
the root systems are isomorphic and the equality follows from a
change of variables; Spiridonov and Vartanov describe these changes
and prove several limiting cases at higher rank
\cite[Section~2]{Spiridonov:2010qv}. The flavored Schur index of $SO(7)$ was
evaluated in \cite[Section~2.2]{Guo:2023mkn}, and the rank-three
unflavored and flavored comparisons are in
\cite[Sections~4.2.2 and~4.3]{Li:2026bch}.

\subsection{Root systems, flavor weights, and the statement}

With $e_1,\ldots,e_n$ the standard basis of $\mathbb R^n$, the root
systems are
\begin{equation}
\begin{split}
 \Phi_{B_n}&=\{\pm e_i:1\leq i\leq n\}
       \cup\{\pm e_i\pm e_j:1\leq i<j\leq n\},\\
 \Phi_{C_n}&=\{\pm2e_i:1\leq i\leq n\}
       \cup\{\pm e_i\pm e_j:1\leq i<j\leq n\},
\end{split}
\label{eq:bc-roots}
\end{equation}
with independent signs. They share the roots $\pm e_i\pm e_j$ and
differ in the roots $\pm e_i$ versus $\pm2e_i$. Both Weyl groups are
generated by permutations and sign changes and have order $2^nn!$.

Let $s$ be the Schur expansion variable, so that the infinite
products have nome $s^2$, and let $x$ be the $SU(2)$ flavor fugacity.
The two adjoint chiral multiplets in the hypermultiplet have weights
\begin{equation}
 u=sx,\qquad v=s/x,\qquad uv=s^2,
 \label{eq:bc-flavor-weights}
\end{equation}
and the unflavored specialization is $x=1$. We initially take
\begin{equation}
 0<|s|<1,\qquad |s|<|x|<|s|^{-1},
 \qquad\text{equivalently}\quad |s|^2<|u|<1,
 \label{eq:bc-domain}
\end{equation}
so that $|u|,|v|<1$ and the denominators of the integrand do not
vanish on the unit torus. The contribution of a root, and of a pair
of opposite roots, is
\begin{equation}
 W_{s,x}(z)=\frac{(z;s^2)_\infty(s^2z;s^2)_\infty}
                    {(uz;s^2)_\infty(vz;s^2)_\infty},
 \qquad R_{s,x}(z)=W_{s,x}(z)W_{s,x}(z^{-1}),
 \label{eq:bc-root-weights}
\end{equation}
and the vacuum-normalized Schur index is
\begin{equation}
 \mathcal I_\Phi(s,x)=\mathcal C_n(s,x)
 \operatorname{CT}_{\boldsymbol z}\prod_{\alpha\in\Phi}W_{s,x}(z^\alpha),
 \qquad
 \mathcal C_n(s,x)=\frac1{2^n n!}
 \left[\frac{(s^2;s^2)_\infty^2}{\theta(u;s^2)}\right]^n,
 \label{eq:bc-index-root}
\end{equation}
where $\theta(u;s^2)=(u;s^2)_\infty(v;s^2)_\infty$ since $uv=s^2$,
and the $n$th power accounts for the $n$ zero weights of the adjoint.
Separating the roots common to both systems,
\begin{equation}
 \mathcal V_n(\boldsymbol z;s,x)
 =\prod_{i<j}R_{s,x}(z_i z_j)R_{s,x}(z_i/z_j),
 \label{eq:bc-common-factors}
\end{equation}
we have
\begin{align}
 \mathcal I_{B_n}(s,x)
 &=\mathcal C_n(s,x)\int_{\mathbb T^n}
 \mathcal V_n(\boldsymbol z;s,x)\prod_iR_{s,x}(z_i)
 \prod_i\frac{\dd z_i}{2\pi\mathrm i z_i},\label{eq:bc-B-index}\\
 \mathcal I_{C_n}(s,x)
 &=\mathcal C_n(s,x)\int_{\mathbb T^n}
 \mathcal V_n(\boldsymbol z;s,x)\prod_iR_{s,x}(z_i^2)
 \prod_i\frac{\dd z_i}{2\pi\mathrm i z_i},
 \label{eq:bc-C-index}
\end{align}
where $\mathbb T$ is the counterclockwise unit circle. The theorem
below thus asserts that replacing each $R_{s,x}(z_i)$ by
$R_{s,x}(z_i^2)$ does not change the integral.

\begin{theorem}\label{thm:bc-flavored}
For every integer $n\geq1$ and parameters in \eqref{eq:bc-domain},
\begin{equation}
 \mathcal I_{B_n}(s,x)=\mathcal I_{C_n}(s,x).
 \label{eq:bc-allrank-theorem}
\end{equation}
The equality extends to $s=0$ for each fixed $x\neq0$, holds
coefficient by coefficient as a power series in $s$ with Laurent
polynomial coefficients in $x$, and persists wherever the two
integrals are defined by meromorphic continuation.
\end{theorem}

The only non-elementary input to the proof is
Theorem~\ref{thm:tools-rains-trace}.

\subsection{An auxiliary gamma integral and Rains's transformation}

Introduce
\begin{equation}
 p_*=u,\qquad q_*=s^2,\qquad
 \Gamma_*(z)=\Gamma(z;u,s^2),\qquad
 G_*(b;z)=\Gamma_*(bz)\Gamma_*(b/z).
 \label{eq:bc-auxiliary-gamma}
\end{equation}
Since $p_*=u$, multiplying the argument of $\Gamma_*$ by $u$ is a
shift by a nome, and the shift identity produces theta functions with
nome $s^2$, the nome of the Schur integrand; the relation $uv=s^2$
supplies the reflection cancellations. The auxiliary nomes need not
equal the two nomes of the full index, since the Schur integral
\eqref{eq:bc-index-root} has already lost its dependence on one of
them. We first take
$s,u$ real with $0<s<1$ and $s^2<u<1$, and choose positive square
roots of positive numbers; complex parameters are treated at the end
of the proof.

The eight boundary parameters come in four pairs $(d_r,s^2/d_r)$, and
the integrand is unchanged if $d_r$ is exchanged with $s^2/d_r$; we
call $d_r$ a representative of its pair. For
$\boldsymbol d=(d_0,d_1,d_2,d_3)$ with nonzero entries, define
\begin{equation}
\begin{split}
 \mathcal J_n(\boldsymbol d;s,u)
 =\mathcal D_n(s,u)\int
 &\prod_{i<j}
 \frac{\Gamma_*(u z_i^{\pm1}z_j^{\pm1})^2
       \Gamma_*(v z_i^{\pm1}z_j^{\pm1})}
      {\Gamma_*(z_i^{\pm1}z_j^{\pm1})}\\
 {}\times&\prod_i\left[
 \frac{G_*(u;z_i^2)}{G_*(1;z_i^2)}
 \prod_{r=0}^3G_*(d_r;z_i)G_*(s^2/d_r;z_i)\right]
 \prod_i\frac{\dd z_i}{2\pi\mathrm i z_i},
\end{split}
\label{eq:bc-J-definition}
\end{equation}
where $v=s^2/u$ and
\begin{equation}
 \mathcal D_n(s,u)=\frac1{2^n n!}
 \left[\kappa(u,s^2)\Gamma_*(v)\Gamma_*(u)^2\right]^n.
 \label{eq:bc-D-normalization}
\end{equation}
Theorem~\ref{thm:tools-rains-trace} gives
\begin{equation}
 \mathcal J_n(\boldsymbol d;s,u)
 =\mathcal J_n(\boldsymbol d';s,u),\qquad
 d_r'=\frac{s^2d_r}{\sqrt{d_0d_1d_2d_3}},
 \label{eq:bc-J-transformation}
\end{equation}
as an identity of integrals over separating contours, continued
meromorphically. Below we work on a domain where both contours are
the unit torus.

To derive \eqref{eq:bc-J-transformation}, including its constant, we
write Theorem~\ref{thm:tools-rains-trace} with Selberg parameter
$\tau$ and boundary variables $\eta_r$. At the parameter value we
need, the kernel is the product
\cite[Proposition~2.10 and Theorem~3.1]{MR3772016}
\begin{equation}
 \mathcal K_{\sqrt{pq/\tau}}^{(n)}
 (\boldsymbol z;\boldsymbol w;\tau;p,q)
 =\prod_{i,j=1}^n
 \Gamma(\sqrt{pq/\tau}\,z_i^{\pm1}w_j^{\pm1};p,q),
 \label{eq:bc-kernel-specialization}
\end{equation}
where $p,q$ are the nomes of the transformation, to be set equal to
$p_*,q_*$. The theorem integrates the diagonal value
$\mathcal K_c^{(n)}(\boldsymbol z;\boldsymbol z;\tau;p,q)$ against the
Selberg density with boundary parameters
$\sqrt{pq/c}\,\eta_r^{\pm1}$, which has pair factor
$\Gamma(\tau z_i^{\pm1}z_j^{\pm1})/
\Gamma(z_i^{\pm1}z_j^{\pm1})$, one-variable denominator
$\Gamma(z_i^{\pm2})$, and prefactor
$\kappa(p,q)^n\Gamma(\tau;p,q)^n/(2^n n!)$. Set
\begin{equation}
 (p,q,\tau,c)=(u,s^2,v,u),\qquad \eta_r=d_r/s.
 \label{eq:bc-trace-parameters}
\end{equation}
Then $c^2=pq/\tau$ follows from $uv=s^2$, and $\sqrt{pq/c}=s$, so the
boundary parameters are $(d_r,s^2/d_r)$. For each $i<j$, the ordered
pairs $(i,j)$ and $(j,i)$ in the kernel give two factors with weight
$u$, and the density gives the factor with weight $v$. For $i=j$, the
four sign choices give $z_i^2$, $1$, $1$, and $z_i^{-2}$, so the
kernel contributes
\begin{equation}
 \Gamma_*(u z_i^{\pm1}z_i^{\pm1})
 =\Gamma_*(u)^2G_*(u;z_i^2),
 \label{eq:bc-kernel-diagonal}
\end{equation}
including two constant factors. Together with the density this gives
\eqref{eq:bc-J-definition}--\eqref{eq:bc-D-normalization}, and the
transformation of $\boldsymbol\eta$ gives
\eqref{eq:bc-J-transformation}.

\subsection{Theta cancellation and the two root systems}

We now reduce $\mathcal J_n$ to the Schur integrals. Put
\begin{equation}
 \Theta_b(z)=\theta(bz;s^2)\theta(b/z;s^2).
 \label{eq:bc-Theta}
\end{equation}
The reflection and shift identities give
\begin{equation}
 G_*(b;z)G_*(us^2/b;z)=1,\qquad
 G_*(ub;z)=\Theta_b(z)G_*(b;z),
 \label{eq:bc-G-shift}
\end{equation}
and hence
\begin{equation}
 G_*(b;z)G_*(s^2/b;z)
 =\frac{G_*(b;z)}{G_*(ub;z)}=\frac1{\Theta_b(z)}.
 \label{eq:bc-boundary-pair}
\end{equation}
For $w=z_iz_j$ or $w=z_i/z_j$, reflection gives
$G_*(v;w)=1/G_*(u^2;w)$, since $us^2/v=u^2$, and two applications of
the shift identity give
\begin{equation}
 \frac{G_*(u;w)^2G_*(v;w)}{G_*(1;w)}
 =\frac{G_*(u;w)^2}{G_*(u^2;w)G_*(1;w)}
 =\frac{\Theta_1(w)}{\Theta_u(w)}=R_{s,x}(w),
 \label{eq:bc-pair-reduction}
\end{equation}
where the last step uses $v=s^2/u$. This reproduces the common factor
$\mathcal V_n$. The one-variable factor is
\begin{equation}
 \frac{G_*(u;z^2)}{G_*(1;z^2)}
 \prod_{r=0}^3G_*(d_r;z)G_*(s^2/d_r;z)
 =\frac{\Theta_1(z^2)}{\prod_{r=0}^3\Theta_{d_r}(z)}.
 \label{eq:bc-one-variable}
\end{equation}

It remains to choose the representatives so that this factor is
$R_{s,x}(z)$ on one side and $R_{s,x}(z^2)$ on the other. The
elementary identities
\begin{equation}
 \theta(y;s^2)\theta(-y;s^2)=\theta(y^2;s^4),\qquad
 \theta(y;s^2)=\theta(y;s^4)\theta(s^2y;s^4)
 \label{eq:bc-theta-elementary}
\end{equation}
give the duplication formula
\begin{equation}
 \Theta_b(z^2)=
 \Theta_{\sqrt b}(z)\Theta_{-\sqrt b}(z)
 \Theta_{s\sqrt b}(z)\Theta_{-s\sqrt b}(z),
 \label{eq:bc-theta-duplication}
\end{equation}
and $\theta(s^2/y;s^2)=\theta(y;s^2)$ gives
\begin{equation}
 \Theta_{s^2/b}(z)=\Theta_b(z).
 \label{eq:bc-complement}
\end{equation}

Choose
\begin{equation}
 \boldsymbol d_B=(-1,s,-s,u).
 \label{eq:bc-dB}
\end{equation}
Its product is $s^2u$, so \eqref{eq:bc-J-transformation} maps it to
\begin{equation}
 \boldsymbol d'
 =(-s/\sqrt u,\ s^2/\sqrt u,\ -s^2/\sqrt u,\ s\sqrt u).
 \label{eq:bc-dprime}
\end{equation}
Replacing the first three entries by their partners $s^2/d_r$, which
changes neither the gamma pairs nor, by \eqref{eq:bc-complement}, the
theta factors, gives
\begin{equation}
 \boldsymbol d_C=(-s\sqrt u,\sqrt u,-\sqrt u,s\sqrt u).
 \label{eq:bc-dC}
\end{equation}
The duplication formula \eqref{eq:bc-theta-duplication} with $b=1$
and with $b=u$ then gives
\begin{align}
 \frac{\Theta_1(z^2)}
 {\Theta_{-1}(z)\Theta_s(z)\Theta_{-s}(z)\Theta_u(z)}
 &=\frac{\Theta_1(z)}{\Theta_u(z)}=R_{s,x}(z),
 \label{eq:bc-singleton-B}\\
 \frac{\Theta_1(z^2)}
 {\Theta_{\sqrt u}(z)\Theta_{-\sqrt u}(z)
  \Theta_{s\sqrt u}(z)\Theta_{-s\sqrt u}(z)}
 &=\frac{\Theta_1(z^2)}{\Theta_u(z^2)}=R_{s,x}(z^2).
 \label{eq:bc-singleton-C}
\end{align}
With the common factor $\mathcal V_n$, these are the integrands of
\eqref{eq:bc-B-index} and \eqref{eq:bc-C-index}.

\subsection{Normalization and a regulated contour limit}

Two points remain: the normalization and the contour. For the first,
reflection and shift give
\begin{equation}
 \Gamma_*(v)=\frac1{\Gamma_*(u^2)},\qquad
 \Gamma_*(u^2)=\theta(u;s^2)\Gamma_*(u),\qquad
 \Gamma_*(u)=\frac{(s^2;s^2)_\infty}{(u;u)_\infty},
 \label{eq:bc-scalar-gamma}
\end{equation}
where the last identity follows from \eqref{eq:tools-gamma} by
telescoping in the first product index. Hence
\begin{equation}
 \kappa(u,s^2)\Gamma_*(v)\Gamma_*(u)^2
 =\frac{(s^2;s^2)_\infty^2}{\theta(u;s^2)},\qquad
 \mathcal D_n(s,u)=\mathcal C_n(s,x).
 \label{eq:bc-normalization-check}
\end{equation}
Second, the representative $-1$ in \eqref{eq:bc-dB} places a pole of
$G_*(-1;z)$ on the unit circle, which is cancelled only by the vector
factor. We therefore replace it by $-\lambda$ with $s^2<\lambda<1$,
keeping $0<s<1$ and $s^2<u<1$. The eight boundary parameters are then
\begin{equation}
 (-\lambda,-s^2/\lambda,s,s,-s,-s,u,v),
 \label{eq:bc-regulated-B}
\end{equation}
and after the transformation \eqref{eq:bc-J-transformation} and the
same exchanges the representatives are
\begin{equation}
 (-s\sqrt{u/\lambda},\sqrt{\lambda u},
  -\sqrt{\lambda u},s\sqrt{u/\lambda}),
 \label{eq:bc-regulated-C}
\end{equation}
each with its partner $s^2/d_r$. All boundary parameters on both
sides lie inside the unit disk; on the transformed side the squared
moduli are $s^2u/\lambda$, $\lambda u$, $s^2\lambda/u$ and
$s^4/(\lambda u)$. Since also $u,v<1$, the unit torus is a separating
contour on both sides of \eqref{eq:bc-J-transformation}.

As $\lambda\to1$, the only singularity reaching the contour comes
from $G_*(-\lambda;z)$ at $z=-1$. Its two approaching simple poles
are compensated by the double zero of the vector factor
$1/G_*(1;z^2)$, whose reciprocal gamma factors at $z^2$ and $z^{-2}$
each vanish at $z=-1$. On $|z|=1$,
\begin{equation}
 \frac{|1-z^2|^2}{|1+\lambda z|^2}
 =\frac{|1-z|^2|1+z|^2}
        {(1-\lambda)^2+\lambda|1+z|^2}
 \leq\frac4\lambda.
 \label{eq:bc-contour-bound}
\end{equation}
All other factors are uniformly bounded on the torus for $\lambda$
near one, and coincident gauge variables give zeros of the vector
factor rather than poles. This gives an integrable bound independent
of $\lambda$, and away from the null sets $z_i=-1$ the integrand
converges to the expression simplified above. By dominated
convergence the left-hand side tends to the $B_n$ integral, with no
residue contribution. The transformed parameters remain inside the
unit disk at $\lambda=1$, so the right-hand side tends to the $C_n$
integral. With \eqref{eq:bc-normalization-check}, this proves
\eqref{eq:bc-allrank-theorem} for real $0<s<1$, $s^2<u<1$.

\subsection{Complex parameters and the unflavored corollary}

For $|s|^2<|u|<1$ the denominators of \eqref{eq:bc-B-index} and
\eqref{eq:bc-C-index} do not vanish on the unit torus and the
products converge uniformly on compact subsets, so both integrals
are holomorphic in $s$ and $u$, with $v=s^2/u$. The identity theorem,
applied first in $s$ at fixed real $0<u<1$ (from $0<s<\sqrt u$ to
$|s|<\sqrt u$) and then in $u$ at fixed complex $s$ (from real
$|s|^2<u<1$ to $|s|^2<|u|<1$), extends the equality to the domain
\eqref{eq:bc-domain}. The square roots of the auxiliary lists do not
appear in the final integrals, so their meromorphic continuations
also agree. For fixed $x\neq0$ both integrals are holomorphic near
$s=0$, and each coefficient of their expansion in $s$ is a Laurent
polynomial in $x$; hence these coefficients agree. This completes the
proof of Theorem~\ref{thm:bc-flavored}.

At $x=1$ we have $u=v=s$ and
\begin{equation}
 W_{s,1}(z)=\frac{(z;s^2)_\infty(s^2z;s^2)_\infty}
 {(sz;s^2)_\infty^2},\qquad
 \mathcal C_n(s,1)=\frac1{2^n n!}
 \left[\frac{(s^2;s^2)_\infty}{(s;s^2)_\infty}\right]^{2n},
 \label{eq:bc-unflavored-corollary}
\end{equation}
which gives a second derivation of the unflavored identity of
\cite{Kulanthaivelu}. The proof uses the Schur relation $uv=s^2$ in
both the reflection and the theta reduction, so it does not extend
directly to the deformed Schur or full elliptic indices; see
Section~\ref{sec:conclusion}.

The equality also fits a conjectural picture based on the vertex
algebras of \cite{Bonetti:2018fqz}, which are labeled by complex reflection
groups: if these are the chiral algebras of the $\mathcal N=4$
theories, the common Weyl group of $B_n$ and $C_n$ implies equal
characters. The type-$A$ results of \cite{Arakawa:2023cki} do not establish this
identification for $B_n$ and $C_n$.

\section{Kutasov--Schwimmer duality}
\label{sec:ks}

The electric theory in Kutasov--Schwimmer duality has gauge group
$SU(N)$, one adjoint chiral multiplet $X$, $F$ fundamental and $F$
antifundamental chiral multiplets, and superpotential
$W=\operatorname{Tr}X^{k+1}$ \cite{Kutasov:1995np}.  Its proposed
magnetic description has gauge group $SU(M)$, where
\begin{equation}
 M=kF-N.
 \label{eq:ks-ranks}
\end{equation}
We take $M\geq1$; for $M=1$ there is no magnetic gauge integral. The
magnetic theory also has $kF^2$ singlets, corresponding to the
electric mesons $\widetilde Q_iX^\ell Q_j$, $0\leq\ell<k$.

The weight $u$ of $X$ satisfies
\begin{equation}
 u^{k+1}=P,
 \label{eq:ks-superpotential}
\end{equation}
and, with $\Gamma(z)=\Gamma(z;p,q)$, $\kappa=\kappa(p,q)$ and the
measure \eqref{eq:tools-measure}, the one-adjoint integral
\eqref{eq:tools-A} is
\begin{align}
 \mathcal A_{N,F}(u;\boldsymbol a,\boldsymbol b;p,q)
 ={}&\Gamma(u)^{N-1}\int \dd\mu_N(\boldsymbol z)
 \prod_{\substack{1\leq i,j\leq N\\i\ne j}}
       \Gamma(uz_i/z_j)\notag\\
 &\hspace{4mm}\times
 \prod_{i=1}^F\prod_{j=1}^N
       \Gamma(a_i z_j,b_i/z_j).
 \label{eq:ks-integral}
\end{align}
For $N=1$, $\mathcal A_{1,F}=\prod_i\Gamma(a_i,b_i)$. The parameters
$a_i,b_i$ are the full chiral weights, including the factors
$P^{R(Q)/2}$ and $P^{R(\widetilde Q)/2}$. We write
\begin{equation}
 A=\prod_{i=1}^F a_i,\qquad B=\prod_{i=1}^F b_i,
 \qquad AB=P^F u^{-2N}.
 \label{eq:ks-balance}
\end{equation}
The last relation is the balancing condition, which expresses the
absence of a gauge anomaly for the $R$-symmetry; for equal
fundamental and antifundamental $R$-charges, $R(X)=2/(k+1)$ and
$R(Q)=1-2N/((k+1)F)$. The conjectured identity is
\begin{align}
 \mathcal A_{N,F}(u;\boldsymbol a,\boldsymbol b)
 \stackrel{?}{=}{}&
 \prod_{i,j=1}^F\prod_{\ell=0}^{k-1}
       \Gamma(u^\ell a_i b_j)
 \mathcal A_{M,F}(u;\widehat{\boldsymbol a},
                            \widehat{\boldsymbol b}),
 \label{eq:ks-proposed}\\
 \widehat a_i={}&\frac{u(A/B)^{k/(2M)}}{a_i},\qquad
 \widehat b_i=\frac{u(B/A)^{k/(2M)}}{b_i}.
 \label{eq:ks-magnetic-parameters}
\end{align}
The prefactor is the contribution of the singlets: a meson at level
$\ell$ and its partners in the magnetic superpotential have weights
with product $u^\ell a_i b_j\,\widehat a_i\widehat b_j
u^{k-1-\ell}=P$. Spiridonov and Vartanov discussed
\eqref{eq:ks-proposed} for general ranks and independent $p,q$
\cite[Sec.~9.1, Eqs.~(9.2)--(9.3)]{Spiridonov:2009za}, and Sudano developed adjoint
deconfinement for index integrals \cite[Appendix~C]{Sudano:2011aa}. Below we
derive two restricted forms of \eqref{eq:ks-proposed} from Rains's
ordinary transformation and its confining case
\cite[Theorem~4.1 and Corollary~4.2]{Rains:2005lkc}: a finite-rank
family at independent nomes, and the restriction to $p=q^k$.

Deconfinement has been used to derive the three-dimensional $U(N)$
adjoint Kim--Park duality and the four-dimensional $USp(2N)$
antisymmetric Intriligator duality \cite{Benvenuti:2024glr}; these results do not
include the four-dimensional $SU(N)$ identity \eqref{eq:ks-proposed}.
The two-node construction of Fang and Kim instead takes
Kutasov--Schwimmer duality as an input to derive an
Intriligator--Leigh--Strassler duality \cite[Sec.~3]{Fang:2026ddf}.

\subsection{An infinite family with independent elliptic nomes}
\label{subsec:ks-independent}

We keep $p,q$ and the flavor parameters generic and fix
\begin{equation}
 F=2,\qquad N=k+1,\qquad M=k-1,\qquad k\geq2,
 \label{eq:ks-special-ranks}
\end{equation}
so that \eqref{eq:ks-balance} becomes $AB=1$ and
$R(Q)=R(\widetilde Q)=0$. In this ultraviolet grading the meson
$\widetilde Q_iQ_j$ has $R=0$, whereas a nonzero gauge-invariant
chiral scalar in a unitary four-dimensional superconformal theory must
have $R\geq2/3$.
Theorem~\ref{thm:ks-independent} is therefore an identity of index
integrals with this grading; relating it to the infrared
superconformal index would require a separate treatment of such
operators and of accidental symmetries (see
\cite{Kutasov:2003iy}).

\begin{theorem}
\label{thm:ks-independent}
Let $k\geq2$, $u^{k+1}=P$, and
$A=a_1a_2=B^{-1}=(b_1b_2)^{-1}$. Then
\begin{align}
 &\mathcal A_{k+1,2}(u;\boldsymbol a,\boldsymbol b)=
 \prod_{i,j=1}^2\Gamma(a_i b_j,u a_i b_j)
 \mathcal A_{k-1,2}\left(
 u;uA^{1/(k-1)}\boldsymbol a,
   uB^{1/(k-1)}\boldsymbol b\right)
 \label{eq:ks-independent-result}
\end{align}
as a meromorphic identity. Multiplication of a vector of parameters by
a scalar acts on each of its entries.
\end{theorem}

This is \eqref{eq:ks-proposed} for these ranks. Since $A/a_1=a_2$ and
$A/a_2=a_1$, the magnetic parameters $uA^{k/(k-1)}/a_i$ are
$uA^{1/(k-1)}a_i$ in reverse order, and similarly for the $b_i$. The
meson factors with $2\leq\ell\leq k-1$ cancel in pairs by reflection:
$(\ell,i,j)$ pairs with $(k+1-\ell,3-i,3-j)$, since
\begin{equation}
 (u^\ell a_i b_j)
 (u^{k+1-\ell}a_{3-i}b_{3-j})=u^{k+1}AB=P,
 \label{eq:ks-meson-reflection}
\end{equation}
and this map is a fixed-point-free involution of that range. Only the
levels $\ell=0,1$ remain.

\subsubsection{The two integral operations}

We use the confining evaluation of Corollary~\ref{cor:tools-confining}
\cite{Rains:2005lkc} in the form
\begin{align}
 &\int \dd\mu_m(\boldsymbol z)
 \prod_{j=1}^m\prod_{\alpha=1}^{m+1}
       \Gamma(v_\alpha z_j,h_\alpha/z_j)=
 \prod_{\alpha,\beta=1}^{m+1}\Gamma(v_\alpha h_\beta)
 \prod_{\alpha=1}^{m+1}
       \Gamma(V/v_\alpha,H/h_\alpha),
 \label{eq:ks-confinement}\\
 &\hspace{20mm}V=\prod_\alpha v_\alpha,\qquad
 H=\prod_\alpha h_\alpha,\qquad VH=P.\notag
\end{align}
We also need a rank-reducing residue limit: for $m\geq2$, a symmetric
meromorphic function $\mathcal F$ and generic parameters,
\begin{align}
 &\lim_{d_0 e\to1}\frac{1}{\Gamma(d_0 e)}
 \int \dd\mu_m(\boldsymbol w)
 \prod_{j=1}^m\Gamma(d_0 w_j,e/w_j)\mathcal F(\boldsymbol w)
 \notag\\
 &\qquad=
 \int \dd\mu_{m-1}(\boldsymbol y)
 \mathcal F(e^{-1/(m-1)}y_1,\ldots,
   e^{-1/(m-1)}y_{m-1},e),
 \label{eq:ks-pinch}
\end{align}
on continued separating contours, provided $\mathcal F$ is regular at
the collision and no other poles collide there; the same holds for a
family of such functions with a locally uniform limit. For the $A_n$
density this is \cite[Lemma~4.5]{Rains:2005lkc}; here $\mathcal F$
also contains an adjoint determinant, whose regularity is checked
below.

To derive \eqref{eq:ks-pinch}, note that the poles at $w_j=e$ and
$w_j=d_0^{-1}$ pinch the contour as $d_0e\to1$. Near $w_j=e$,
\begin{equation}
 \Gamma(e/w_j)\frac{\dd w_j}{w_j}
 \sim\frac{1}{\kappa}\frac{\dd w_j}{w_j-e},
 \qquad
 \lim_{v\to1}(1-v)\Gamma(v)=\kappa^{-1},
 \label{eq:ks-local-residue}
\end{equation}
and the factor $\Gamma(d_0e)$ at this point cancels the prefactor. The
$m$ choices of the variable fixed at $e$, together with the residue
$1/\kappa$, change the normalization by
\begin{equation}
 \frac{\kappa^{m-1}}{m!}\times\frac{m}{\kappa}
 =\frac{\kappa^{m-2}}{(m-1)!}.
 \label{eq:ks-pinch-normalization}
\end{equation}
For the remaining $w_i$, the factors $\Gamma(d_0 w_i,e/w_i)$ at
$d_0=e^{-1}$ cancel the vector factors $\Gamma(w_i/e,e/w_i)$, and
since $\prod_{i<m}w_i=e^{-1}$, the substitution
$w_i=e^{-1/(m-1)}y_i$ restores $\prod_i y_i=1$.

The integral over the contour away from the collision is bounded and
multiplied by $1/\Gamma(d_0 e)\to0$. After one residue, the factors
that could pinch a second variable at $e$ have cancelled against the
vector denominator, so only single residues contribute. This proves
\eqref{eq:ks-pinch} for generic parameters, and meromorphic
continuation gives the other regular values. In our application the
adjoint factors at $w_i=w_j=e$ have argument $u$ and are generically
regular.

\subsubsection{The auxiliary integral and its regulated rank reduction}
\label{subsubsec:ks-regulated-reduction}

Deconfinement replaces an adjoint by fundamentals of an auxiliary
gauge group; Luty, Schmaltz and Terning described the module for an
$SU(N)$ adjoint \cite{Luty:1996cg}, and Sudano proved the
corresponding index identity using the confining integral
\cite[Appendix~C]{Sudano:2011aa}. We use an auxiliary $SU(k)$ node with
$u^{k+1}=P$, integrate in both orders, and evaluate the second order
by a regulated rank reduction.

Choose a branch $c^k=b_2$, put $d=u/c$, and vary only the
regulator $\lambda$, keeping $p,q,u,a_i,b_i,c,d$ fixed. Define
\begin{equation}
 x=\frac{b_1}{\lambda}=\frac{1}{\lambda A c^k},\qquad
 \eta=\lambda u b_2,\qquad E=A c^k,\qquad H=\frac{P}{E}.
 \label{eq:ks-aux-parameters}
\end{equation}
The parameters needed after integrating the $SU(k+1)$ node are
\begin{equation}
\begin{gathered}
 e=A c^{k-1},\qquad d_0=cx=\frac{1}{\lambda e},\qquad
 g=c\eta,\\
 f_i=d a_i\quad(i=1,2),\qquad h=\frac{H}{d}=\frac{u^k}{e}.
\end{gathered}
 \label{eq:ks-residue-parameters}
\end{equation}
Then $cd=u$, $x\eta d^k=H$, $EH=P$, and $d_0e=\lambda^{-1}$.
Table~\ref{tab:ks-regulator-weights} lists the weights in both
integration orders and their values at $\lambda=1$; list operations
are entrywise. Only $x,\eta,d_0,g$ depend on $\lambda$, and no
independent flavor fugacities are introduced.

\begin{table}[!ht]
\centering
\small
\setlength{\tabcolsep}{4pt}
\renewcommand{\arraystretch}{1.3}
\begin{tabular}{@{}llcc@{}}
\hline
Node & Representation & Weights at general $\lambda$ & At $\lambda=1$ \\
\hline
Original $SU(k)$ & fundamental & $c\boldsymbol z$ & $c\boldsymbol z$ \\
 & antifundamental & $d/\boldsymbol z$ & $d/\boldsymbol z$ \\
Original $SU(k+1)$ & fundamental & $(a_1,a_2,c\boldsymbol w)$
 & $(a_1,a_2,c\boldsymbol w)$ \\
 & antifundamental & $(x,\eta,d/\boldsymbol w)$
 & $(b_1,ub_2,d/\boldsymbol w)$ \\
Residual $SU(k)$ & fundamental & $(d_0,g,h)$
 & $(e^{-1},ucb_2,u^k/e)$ \\
 & antifundamental & $(e,f_1,f_2)$ & $(e,f_1,f_2)$ \\
\hline
\end{tabular}
\caption{Weights in the auxiliary double integral and in the residual
$SU(k)$ integral obtained by integrating $\boldsymbol z$ first.
The residual integral also contains one adjoint of fixed weight $u$.
The original two nodes are linked by the weights $c,d$.}
\label{tab:ks-regulator-weights}
\end{table}

Consider the double integral
\begin{align}
 J(\lambda)={}&\frac{1}{\Gamma(u)}
 \iint \dd\mu_{k+1}(\boldsymbol z)\dd\mu_k(\boldsymbol w)
 \prod_{i=1}^{k+1}
       \Gamma(a_1z_i,a_2z_i,x/z_i,\eta/z_i)\notag\\
 &\hspace{12mm}\times
 \prod_{i=1}^{k+1}\prod_{j=1}^k
       \Gamma(cz_iw_j,d/(z_iw_j)).
 \label{eq:ks-double}
\end{align}
In either order of integration, the variables integrated first have
the matter content of \eqref{eq:ks-confinement}, so no adjoint duality
is assumed.

\paragraph{Absolute convergence and the two orders of integration.}
Take $0<u<1$, $0<|p|,|q|<1$, $pq=u^{k+1}$, and initially choose
\begin{equation}
 c=d=u^{1/2},\quad a_1=a_2=u^{1/4},\quad
 b_2=u^{k/2},\quad b_1=u^{-(k+1)/2},\quad
 \lambda_0=u^{-(2k+3)/4}.
 \label{eq:ks-initial-domain}
\end{equation}
Then $x=\eta=u^{1/4}$. On the unit $SU(k+1)\times SU(k)$ torus,
all $z_i,w_j$, including those fixed by the determinant constraints,
have modulus one. The arguments of the matter gamma functions in
\eqref{eq:ks-double} have modulus $u^{1/4}$ or $u^{1/2}$.
They avoid every pole $p^{-m}q^{-n}$, $m,n\geq0$, whose modulus is
at least one, and the products converge uniformly on this compact
set. The vector factors are regular as well: pairing opposite roots
gives
\begin{equation}
 \frac{1}{\Gamma(v)\Gamma(v^{-1})}
 =-v^{-1}\theta(v;p)\theta(v;q),
 \label{eq:ks-vector-regularity}
\end{equation}
so coincident eigenvalues give zeros, not poles, and $1/\Gamma(u)$ is
finite since $|P|<u<1$. The full integrand is thus continuous on
the compact torus $\mathbb T^{2k-1}$, and Fubini's theorem permits
either order of integration.

For fixed base parameters in \eqref{eq:ks-initial-domain}, the same
argument holds throughout the nonempty open annulus
\begin{equation}
 u^{-(k+1)/2}<|\lambda|<u^{-(k+2)/2},
 \label{eq:ks-initial-annulus}
\end{equation}
which is $|x|,|\eta|<1$. Small complex perturbations within the
balancing conditions preserve these strict inequalities, giving an
open parameter domain. At
$\lambda_0$, the remaining
integral after integrating $\boldsymbol w$ has weights
$a_i,x,\eta=u^{1/4}$, $c^k=d^k=u^{k/2}$ and adjoint weight $u$.
After integrating $\boldsymbol z$, its weights are
$d_0=g=f_i=u^{3/4}$, $e=h=u^{k/2}$ and $u$,
so both resulting integrals also converge on their unit tori. The
endpoint $\lambda=1$ is treated below.

First integrate over $\boldsymbol w$. Its $k+1$ pairs of flavor
parameters are $cz_i,d/z_i$, whose total product is
$(cd)^{k+1}=P$. Formula \eqref{eq:ks-confinement} gives
\begin{equation}
 \Gamma(u)^{k+1}
 \prod_{i\ne j}\Gamma(uz_i/z_j)
 \prod_{i=1}^{k+1}\Gamma(c^k/z_i,d^kz_i).
 \label{eq:ks-first-integration}
\end{equation}
The first factor comes from the $i=j$ mesons and the last product from
the baryons. At $\lambda=1$,
\begin{equation}
 x=b_1,\qquad c^k=b_2,\qquad \eta d^k=P.
 \label{eq:ks-electric-endpoint}
\end{equation}
Hence $\Gamma(\eta/z_i,d^kz_i)=1$ by reflection. At regular
endpoint parameters and with continued separating contours,
\begin{equation}
 \lim_{\lambda\to1}J(\lambda)
 =\mathcal A_{k+1,2}(u;\boldsymbol a,\boldsymbol b).
 \label{eq:ks-electric-limit}
\end{equation}
The scalar factor $\Gamma(u)^{k+1}/\Gamma(u)=\Gamma(u)^k$ is the
contribution of the $k$ zero weights of the electric adjoint.

Now integrate over $\boldsymbol z$ first. Its $k+2$ fundamental
parameters and $k+2$ antifundamental parameters are
\begin{equation}
 (a_1,a_2,cw_1,\ldots,cw_k),\qquad
 (x,\eta,d/w_1,\ldots,d/w_k),
 \label{eq:ks-second-flavors}
\end{equation}
whose products are $E$ and $H$, with $EH=P$; the weights of the
residual integral are listed in Table~\ref{tab:ks-regulator-weights}.
The factors independent of $\boldsymbol w$, apart from
$\Gamma(u)^{k-1}$, are
\begin{equation}
 \mathcal C(\lambda)
 =\prod_{i=1}^2\Gamma(a_i x,a_i\eta,c^k a_i)
   \Gamma(H/\eta),
 \label{eq:ks-scalar-C}
\end{equation}
and \eqref{eq:ks-confinement} gives
\begin{align}
 J(\lambda)={}&
 \frac{\Gamma(u)^{k-1}\mathcal C(\lambda)}
      {\Gamma(\lambda^{-1})}
 \int \dd\mu_k(\boldsymbol w)
 \prod_{i\ne j}\Gamma(uw_i/w_j)\notag\\
 &\hspace{4mm}\times
 \prod_{j=1}^k
 \Gamma(d_0 w_j,gw_j,hw_j,e/w_j,f_1/w_j,f_2/w_j).
 \label{eq:ks-second-integration}
\end{align}
Here $\Gamma(e/w_j)=\Gamma(E/(cw_j))$ and
$\Gamma(hw_j)=\Gamma(H/(d/w_j))$ are baryon factors; the remaining
baryons are in $\mathcal C$ and in
\begin{equation}
 \Gamma(H/x)=\Gamma(\lambda P)
 =\frac{1}{\Gamma(\lambda^{-1})},
 \label{eq:ks-zero-factor}
\end{equation}
which tends to zero as $\lambda\to1$ and is compensated by a pole of
the integral.

\paragraph{Continuation and the isolated endpoint pinch.}
The factors $\Gamma(e/w_j)$ and $\Gamma(d_0w_j)$ have pole families
\begin{equation}
 w_j=e p^m q^n,\qquad
 w_j=d_0^{-1}p^{-m'}q^{-n'},\qquad m,n,m',n'\geq0.
 \label{eq:ks-moving-poles}
\end{equation}
A collision requires $d_0e=p^{-(m+m')}q^{-(n+n')}$.
Since $0<|p|,|q|<1$, at $d_0e=1$ this forces
$m=n=m'=n'=0$, so only the leading pair $w_j=e$ and
$w_j=d_0^{-1}=\lambda e$ meets; the contours are continued to
separate the two families until this collision.

The required nonresonance holds on an explicit open domain. Take
$0<u<1$, $pq=u^{k+1}$, $|p|,|q|<u^{1/2}$, and set
\begin{equation}
 \begin{gathered}
 \delta=\frac{k-1}{2k},\qquad
 \epsilon=\frac1{2(k+1)},\qquad
 \alpha=\frac{k-1}{4k(k+1)},\\
 e=u^\delta,\quad c=u^\epsilon,\quad
 a_1=a_2=u^\alpha,\quad b_2=c^k,\quad b_1=(ec)^{-1}.
 \end{gathered}
 \label{eq:ks-regular-endpoint-domain}
\end{equation}
These choices obey $AB=1$ and $e=Ac^{k-1}$. The relevant weights,
with $\lambda_0=u^{-1/2}$ as an initial point, are
\begin{equation}
 \begin{aligned}
 (x,\eta)|_{\lambda_0}
 &=\bigl(u^{1/[2k(k+1)]},u^{1-\epsilon}\bigr),\\
 (g,h,f_i)
 &=\bigl(\lambda u^{3/2},u^{k-\delta},u^{1-1/(4k)}\bigr).
 \end{aligned}
 \label{eq:ks-endpoint-weights}
\end{equation}
All matter weights in the original double integral then have modulus
less than one, so its unit-torus Fubini argument applies.

Write $\zeta=\lambda e=d_0^{-1}$ and continue from
$\zeta_0=u^{-1/(2k)}>1$ towards $e$. Only the leading outer poles
$w_j=\zeta$ cross the unit torus: their shifted partners remain outside
because $\zeta\ge e>u^{1/2}>|p|,|q|$. Small contour deformations
separate the Weyl-equivalent crossings. After the residue at
$w_k=\zeta$, each remaining $\Gamma(w_i/\zeta)$ cancels its vector
denominator, so no second residue of this type arises. On the residual torus
$w_i=\zeta^{-1/(k-1)}y_i$, $|y_i|=1$, $\prod_{i<k}y_i=1$, one has
for $e\le\zeta\le1$
\begin{equation}
 \begin{gathered}
 |uw_i/w_k|=u\zeta^{-k/(k-1)}
 \le ue^{-k/(k-1)}=u^{1/2}<1,\\
 |uw_k/w_i|=u\zeta^{k/(k-1)}\le u<1,\\
 |uw_i/w_j|=u<1\quad(i\ne j<k),\\
 \max\{|gw_i|,|hw_i|,|f_a/w_i|,
             |g\zeta|,|h\zeta|,|f_a/\zeta|\}<1,\\
 |e/w_i|<1,\qquad
 |\zeta/w_i|=\zeta^{k/(k-1)}\in[u^{1/2},1],
 \quad i<k,\ a=1,2.
 \end{gathered}
 \label{eq:ks-residue-contour-bounds}
\end{equation}
At $\zeta=e$, the two cross arguments have moduli $u^{1/2}$ and
$u^{3/2}$. Since every gamma pole has modulus at least one, these
strict bounds exclude further adjoint pinches in a complex
neighborhood of the endpoint. The remaining cross ratio
$\Gamma(e/w_i)/\Gamma(\zeta/w_i)$ has no poles from its denominator,
since $|\zeta/w_i|>|P|$, so the residue integral is regular up to the
endpoint.

Let $K_{\mathrm{cont}}$ and $K_{\mathbb T}$ denote the integral in
\eqref{eq:ks-second-integration}, without scalar prefactors, on the
continued and fixed unit contours. The single residues, normalized as
in \eqref{eq:ks-pinch-normalization}, give
\begin{equation}
 \begin{aligned}
 K_{\mathrm{cont}}(\lambda)
 &=K_{\mathbb T}(\lambda)
   +\Gamma(\lambda^{-1})\mathcal R(\lambda),\\
 \lim_{\lambda\to1}
 \frac{K_{\mathrm{cont}}(\lambda)}{\Gamma(\lambda^{-1})}
 &=\mathcal R(1).
 \end{aligned}
 \label{eq:ks-normalized-local-limit}
\end{equation}
Here $K_{\mathbb T}$ is bounded near one; the bounds above and the
cross ratios tending to one give the locally uniform residue limit
$\mathcal R(1)$ in \eqref{eq:ks-pinch}. Earlier scalar poles are
bypassed, $\mathcal C$ is regular near the endpoint, and these
properties persist in an open neighborhood within the balancing
variety, from which the identity continues meromorphically.

\paragraph{Identification of the magnetic integral.}

Apply \eqref{eq:ks-pinch} to \eqref{eq:ks-second-integration}: one
variable is fixed at $w_k=e$ and the others are rescaled to have
product one. The adjoint factors involving $w_k$,
\begin{equation}
 \prod_{j<k}\Gamma(uw_j/e,ue/w_j),
 \label{eq:ks-adjoint-cross}
\end{equation}
act as an additional fundamental of weight $u/e$ and antifundamental
of weight $ue$. At the endpoint,
\begin{equation}
 h=\frac{u^k}{e},\qquad (ue)h=u^{k+1}=P,
 \label{eq:ks-cross-cancellation}
\end{equation}
so the second of these factors cancels $\Gamma(hw_j)$. The remaining
fundamental weights are $g,u/e$ and the antifundamental weights are
$f_1,f_2$.

The scalar left by the residue is
\begin{equation}
 \mathcal C(1)\Gamma(u)\Gamma(ge,he)
       \prod_{i=1}^2\Gamma(f_i/e).
 \label{eq:ks-residue-scalar}
\end{equation}
The extra $\Gamma(u)$ accounts for the difference between the $k-1$
zero weights in the prefactor of \eqref{eq:ks-second-integration} and
the $k-2$ of the $SU(k-1)$ integral. The scalar factors simplify by
\begin{equation}
 he=u^k,\qquad
 ge=u\frac{b_2}{b_1},\qquad
 \frac{H}{\eta}=u^k\frac{b_1}{b_2},\qquad
 \frac{f_i}{e}=u a_i b_1.
 \label{eq:ks-scalar-identities}
\end{equation}
The first identity cancels $\Gamma(u)\Gamma(he)$, and the middle two
cancel $\Gamma(ge)\Gamma(H/\eta)$. The remaining factors in
$\mathcal C(1)$ are $\Gamma(a_i b_1,u a_i b_2,a_i b_2)$, and with
$\Gamma(f_i/e)$ they give
\begin{equation}
 \prod_{i,j=1}^2\Gamma(a_i b_j,u a_i b_j).
 \label{eq:ks-final-mesons}
\end{equation}

Finally set $w_j=e^{-1/(k-1)}y_j$, which multiplies fundamental
weights by $e^{-1/(k-1)}$ and antifundamental weights by
$e^{1/(k-1)}$. Using $c^k=b_2$ and $A=B^{-1}$,
\begin{align}
 ge^{-1/(k-1)}&=uB^{1/(k-1)}b_2,
 & (u/e)e^{-1/(k-1)}&=uB^{1/(k-1)}b_1,\notag\\
 f_i e^{1/(k-1)}&=uA^{1/(k-1)}a_i,
 \label{eq:ks-final-flavors}
\end{align}
and replacing every $y_j$ by $y_j^{-1}$, which exchanges fundamentals
and antifundamentals, gives the integral in
\eqref{eq:ks-independent-result}. This proves the theorem. Different
consistent choices of the roots are absorbed by multiplying the
magnetic integration variables by a central element of $SU(k-1)$.

For $k=2$, $u^3=P$ and $M=1$, and the theorem gives the evaluation
\begin{equation}
 \mathcal A_{3,2}(u;\boldsymbol a,\boldsymbol b)
 =\prod_{i,j=1}^2\Gamma(a_i b_j,u a_i b_j)
   \prod_{i=1}^2\Gamma(uA a_i,uB b_i).
 \label{eq:ks-SU3-example}
\end{equation}
The two meson levels and the dressed baryons are the confined spectrum
of the cubic $SU(2F-1)$ theory at $F=2$
\cite[Sec.~2]{Csaki:1998fm}. For $k=3$ the theorem relates
$SU(4)$ to $SU(2)$.

\subsection{Reduction on the locus \texorpdfstring{$p=q^k$}{p equals q to the k}}
\label{subsec:ks-commensurate}

Take positive integers $N,F,k$ with $M=kF-N\geq1$. Set
\begin{equation}
 p=q^k=:\rho,\qquad u=q,\qquad 0<|q|<1.
 \label{eq:ks-nome-locus}
\end{equation}
Here $P=pq=q^{k+1}$. Maruyoshi, Nardoni and Song used the same
specialization for adjoint $SU(N)$ SQCD, with $2\leq k<N$ and
$\gcd(k,N)=1$, to match its index to a gauged $\mathcal D_k[SU(N)]$
description \cite[Sec.~3.3]{Maruyoshi:2023mnv}. The
balancing condition becomes
\begin{equation}
 AB=q^{(k+1)F-2N}.
 \label{eq:ks-commensurate-balance}
\end{equation}
We reduce both sides of \eqref{eq:ks-proposed} to the ordinary SQCD
transformation at nomes $(\rho,\rho)$. The resulting identity holds on
each component of the balanced parameter cover on which the electric
integral, the magnetic integral and the meson prefactor are
simultaneously regular on a nonempty relatively open set. This
condition is not automatic, since the balancing condition can force a
meson pole for every choice of flavor fugacities; we give a sufficient
condition and a singular example at the end of this subsection.

\subsubsection{Three product identities}

The first is a special case of the multiplication formulas for the
elliptic gamma function \cite{MR2184012},
\begin{equation}
 \Gamma(z;\rho,q)
 =\prod_{\alpha=0}^{k-1}\Gamma(q^\alpha z;\rho,\rho),
 \label{eq:ks-gamma-splitting}
\end{equation}
obtained by writing $n=kj+\alpha$ in the denominator factors
$1-z\rho^m q^n$ of \eqref{eq:tools-gamma} and $n=kj+k-1-\alpha$ in the
numerator factors $1-\rho^{m+1}q^{n+1}/z$.

Second, the shift and reflection identities give
\begin{align}
 \frac{\Gamma(qx,q/x;\rho,q)}
      {\Gamma(x,1/x;\rho,q)}
 =\theta(x;\rho)\theta(1/x;\rho)=\frac{1}{\Gamma(x,1/x;\rho,\rho)};
 \label{eq:ks-root-splitting}
\end{align}
for the second equality, apply the shift identity twice at equal
nomes,
\[
\Gamma(\rho^2x;\rho,\rho)
=\theta(\rho x;\rho)\theta(x;\rho)
\Gamma(x;\rho,\rho),
\]
and use $\theta(\rho x;\rho)=\theta(1/x;\rho)$ and reflection. Thus,
for each pair of gauge variables, the adjoint and vector factors at
nomes $(\rho,q)$ combine into the vector factors at nomes
$(\rho,\rho)$. Third,
\begin{equation}
 \Gamma(q;\rho,q)=
 \frac{(\rho;\rho)_\infty}{(q;q)_\infty},\qquad
 (\rho;\rho)_\infty(q;q)_\infty\Gamma(q;\rho,q)
 =(\rho;\rho)_\infty^2,
 \label{eq:ks-normalization-splitting}
\end{equation}
where the first identity follows from the limit $z\to1$ of
$\Gamma(qz;\rho,q)=\theta(z;\rho)\Gamma(z;\rho,q)$, using
$\theta(z;\rho)/(1-z)\to(\rho;\rho)_\infty^2$ and
\eqref{eq:tools-residue}.

\subsubsection{Reduction to the ordinary SQCD theorem}

Apply \eqref{eq:ks-gamma-splitting} to every fundamental factor in
the electric integral, and apply
\eqref{eq:ks-root-splitting}--\eqref{eq:ks-normalization-splitting}
to its gauge factors. The result is the ordinary $SU(N)$ SQCD integral
at nomes $(\rho,\rho)$ with $kF$ flavors, which occur in strings of
length $k$:
\begin{equation}
 a_{i,\alpha}=q^\alpha a_i,\qquad
 b_{i,\alpha}=q^\alpha b_i,
 \qquad 1\leq i\leq F,\quad0\leq\alpha<k.
 \label{eq:ks-flavor-strings}
\end{equation}
Their products are
\begin{equation}
 A_{\rm str}=q^{Fk(k-1)/2}A^k,\qquad
 B_{\rm str}=q^{Fk(k-1)/2}B^k,
 \label{eq:ks-string-products}
\end{equation}
so that
\begin{equation}
 A_{\rm str}B_{\rm str}
 =q^{Fk(k-1)+k((k+1)F-2N)}
 =q^{2kM}=\rho^{2M}.
 \label{eq:ks-SQCD-balance}
\end{equation}
This is the balancing condition \eqref{eq:tools-balance} for $kF$
flavors at nomes $(\rho,\rho)$, with $M=kF-N$, so Rains's
transformation \cite[Theorem~4.1]{Rains:2005lkc} applies on the
regular components specified above. Its meson factor becomes
\begin{align}
 &\prod_{i,j=1}^F\prod_{\alpha,\beta=0}^{k-1}
       \Gamma(q^{\alpha+\beta}a_i b_j;\rho,\rho)=\prod_{i,j=1}^F\prod_{\alpha=0}^{k-1}
       \Gamma(q^\alpha a_i b_j;\rho,q),
 \label{eq:ks-string-mesons}
\end{align}
by \eqref{eq:ks-gamma-splitting} applied to the product over $\beta$,
which is the Kutasov--Schwimmer meson factor. The magnetic parameters
of the SQCD transformation are
\begin{equation}
 \frac{A_{\rm str}^{1/M}}{q^\alpha a_i},\qquad
 \frac{B_{\rm str}^{1/M}}{q^\alpha b_i}.
 \label{eq:ks-magnetic-strings-first}
\end{equation}
With $\beta=k-1-\alpha$,
\[
\frac{A_{\rm str}^{1/M}}{q^\alpha a_i}
=q^\beta\frac{q^{-(k-1)}A_{\rm str}^{1/M}}{a_i},
\]
and using $kF=N+M$ in \eqref{eq:ks-string-products} gives
\begin{equation}
 q^{-(k-1)}A_{\rm str}^{1/M}
 =q^{(k-1)(N-M)/(2M)}A^{k/M}
 =q(A/B)^{k/(2M)}.
 \label{eq:ks-magnetic-string-match}
\end{equation}
For the last equality, substitute $B=q^{(k+1)F-2N}/A$; the exponent
of $q$ on the right becomes $1-k((k+1)F-2N)/(2M)$, which equals
$(k-1)(N-M)/(2M)$ by $M=kF-N$. With $A$ and $B$ exchanged the same
holds for the second string, so \eqref{eq:ks-magnetic-strings-first}
is
\begin{equation}
 q^\beta\widehat a_i,\qquad q^\beta\widehat b_i,
 \qquad 0\leq\beta<k,
 \label{eq:ks-magnetic-strings-final}
\end{equation}
with $\widehat a_i,\widehat b_i$ as in
\eqref{eq:ks-magnetic-parameters} and $u=q$.

Applying the three product identities in reverse to the magnetic SQCD
integral reconstructs its adjoint, its normalization and its $F$
flavors at nomes $(\rho,q)$. Together with \eqref{eq:ks-string-mesons},
this gives, on each generically regular component of the balanced
parameter space,
\begin{equation}
 \boxed{
 \begin{aligned}
 &\mathcal A_{N,F}(q;\boldsymbol a,\boldsymbol b;\rho,q)\\
 &\quad=
 \prod_{i,j=1}^F\prod_{\ell=0}^{k-1}
       \Gamma(q^\ell a_i b_j;\rho,q)
 \mathcal A_{M,F}(q;\widehat{\boldsymbol a},
                    \widehat{\boldsymbol b};\rho,q),
 \qquad \rho=q^k,
 \end{aligned}}
 \label{eq:ks-commensurate-result}
\end{equation}
with compatible fractional powers. Rains's theorem at equal nomes
follows from the general case by meromorphic continuation, and outside
a unit-torus chamber \eqref{eq:ks-commensurate-result} is understood
with the continued contours.

A regular chamber exists, for example, when $|N-M|<F$: take $0<q<1$,
choose the positive roots and put
\[
 a_i=b_i=q^{(F+M-N)/(2F)}.
\]
The balance holds, and every magnetic flavor has weight
$q^{(F+N-M)/(2F)}$. Both exponents are positive, so the electric and
magnetic integrals converge on the unit tori and the meson factors
have no poles. These properties persist on an open neighborhood in
the balanced locus.

For a singular family, take $F=1$ and $d=N-M>0$. The balance forces
$a_1b_1=q^{1-d}$, and the $k$ meson arguments have exponents
$1-d,\ldots,k-d$. Exactly $d$ of these are nonpositive and give gamma
poles. The remaining exponents lie below $k+1$, the first gamma zero
in this sequence of powers of $q$. At $0<q<1$ and
$a_1=b_1=q^{(1-d)/2}$ the magnetic weights both equal $q^{(1+d)/2}$.
Its unit-torus density is positive away from coincident eigenvalues,
since opposite weights occur in complex-conjugate pairs; thus the
magnetic integral is generically nonzero. For instance, $k=3$, $N=2$,
$M=1$ forces $a_1b_1=1$. The magnetic expression contains $\Gamma(1)$,
while the other factors, including the two $SU(1)$ chiral factors,
are generically finite and nonzero. Meromorphic continuation does not
assign a value on such a polar locus, and an identity there would
require a specified transverse limit or residue.

Both nomes in \eqref{eq:ks-commensurate-result} are nonzero; the
$p\to0$ degenerations of s-confining indices studied by Do\u{g}an and
Gahramanov give three-dimensional half-indices
\cite{Dogan:2026alu}.

By the symmetry $p\leftrightarrow q$ the same result holds on
$q=p^k$. Neither this result nor Theorem~\ref{thm:ks-independent}
implies the unrestricted identity \eqref{eq:ks-proposed}.

\section{Argyres--Douglas quiver indices}
\label{sec:ad-quivers}

The single-group Lagrangians for the $(A_1,A_m)$ and $(A_1,D_m)$
Argyres--Douglas theories, with gauge group $USp(2n)$ or $SU(n+1)$,
are due, respectively, to Maruyoshi and Song \cite{Maruyoshi:2016aim} and
to Agarwal, Maruyoshi and Song \cite{Agarwal:2016pjo}. The $SU(2)^n$ quivers of Cho and Song
\cite{Cho:2026tqf} have one adjoint at each node, fundamentals at one or
both ends, bifundamentals between neighboring nodes, and gauge
singlets coupled to selected gauge-invariant operators. Lee, Razamat,
Song and Waddleton give a geometric construction of the
$D_p(\mathfrak{su}_2)$ quivers \cite[Section~2]{Lee:2026jxa}.

We prove the full flavored identity for the rank-one $H_1$ theory by
specializing van de Bult's transformation
\cite[Theorem~3.2]{MR2787288}, the case $n=1$ of
\cite[Theorem~3.22]{MR3772016}. For all four families and every
$n$, we then evaluate both descriptions in the generalized Coulomb
limit $q\to0$ with $p$ and $\xi=pq/t$ fixed, using the Macdonald
constant-term theorem. The result, Theorem~\ref{thm:ad-two-parameter},
is the generalized Coulomb index conjectured in
\cite[Conjecture~4]{Buican:2023efi}; it is independent of the flavor
fugacities and reduces at $p=0$ to the Coulomb branch generating
function. The comparison at full or Schur fugacities for $n\geq2$
remains open.

\subsection{Two full-index formulas for \texorpdfstring{$H_1$}{H1}}

The rank-one theory $H_1=(A_1,A_3)=(A_1,D_3)$ has two ultraviolet
Lagrangians, with two fundamentals \cite{Maruyoshi:2016aim} and with four
fundamentals \cite{Agarwal:2016pjo}, which are also the $n=1$ members of the
Cho--Song families \cite{Cho:2026tqf}. Agarwal, Maruyoshi and Song
compared the flavored index expansions
\cite[Section~5.2, Eqs.~(5.18)--(5.19)]{Agarwal:2016pjo}, and Cho, Maruyoshi,
Nardoni and Song compared the two superpotentials, matched their
moment maps and reported agreement of the indices
\cite[Section~4.1]{Cho:2024civ}. We prove the equality with $p,q,t$
and the flavor fugacity independent.

Set
\begin{equation}
 u=(P/t)^{1/3}.
\end{equation}
After placing the common singlet and zero-weight contributions outside
the integrals, define
\begin{align}
 F_2(x)
 &=\frac{\kappa}{2}\oint\frac{\dd z}{2\pi\mathrm i z}
   \frac{\Gamma(uz^{\pm2})}{\Gamma(z^{\pm2})}
   \Gamma(P^{1/2}u^{-2}x^{\pm1}z^{\pm1}),
 \label{eq:ad-h1-two}\\
 F_4(y)
 &=\frac{\kappa}{2}\oint\frac{\dd z}{2\pi\mathrm i z}
   \frac{\Gamma(uz^{\pm2})}{\Gamma(z^{\pm2})}
   \Gamma(P^{1/2}u^{-5/2}z^{\pm1},
          P^{1/2}u^{-1/2}z^{\pm1},
          P^{1/2}u^{-1/2}y^{\pm1}z^{\pm1}).
 \label{eq:ad-h1-four}
\end{align}
We will prove
\begin{equation}
 \boxed{F_2(x)=F_4(x^2).}
 \label{eq:ad-h1-identity}
\end{equation}
The interacting indices are
\begin{equation}
 \mathcal I_{H_1}^{(2)}(x)
 =\frac{\Gamma(u^4)\Gamma(u)}{\Gamma(u^2)}F_2(x),\qquad
 \mathcal I_{H_1}^{(4)}(y)
 =\frac{\Gamma(u^4)\Gamma(u)}{\Gamma(u^2)}F_4(y).
 \label{eq:ad-h1-prefactor}
\end{equation}
In the four-fundamental description the singlet $M_1$, which is free
at $n=1$, has been removed, as explained before
\cite[Eq.~(3.12)]{Cho:2026tqf}; it would contribute $\Gamma(u^3)$. The
flavor map $y=x^2$ matches the moment maps: their charged components
have weights $x^{\pm2}$ in the two-fundamental description, whose
fundamentals have weights $x^{\pm1}$, and $y^{\pm1}$ in the
four-fundamental description.

\subsection{Specialization of the rank-one transformation}

The same transformation proves crossing symmetry of the $SU(2)$ index
with four flavors \cite[Eqs.~(3.17)--(3.19)]{Gadde:2009kb}.
We apply Corollary~\ref{thm:tools-bult}, with $E$ and $\mathcal N_b$
as in \eqref{eq:tools-bult-normalization}--\eqref{eq:tools-bult-E},
and
\begin{equation}
 b=c=u,\qquad t_r=h_r,\qquad
 \boldsymbol h=P^{1/2}(xu^{-2},xu^{-1},x,x^{-1}u^{-2}).
 \label{eq:ad-h1-bult-parameters}
\end{equation}
Since $h_1h_2h_3h_4=P^2x^2u^{-5}$, the rescaling in
\eqref{eq:tools-bult-transformation} is
\begin{equation}
 \begin{aligned}
 h_r&\longmapsto h'_r=\eta h_r,\qquad \eta=u^{3/2}/x,\\
 \boldsymbol h'
 &=P^{1/2}(u^{-1/2},u^{1/2},u^{3/2},x^{-2}u^{-1/2}),
 \end{aligned}
 \label{eq:ad-bult-map}
\end{equation}
with consistent choices of powers. By the invariance of $\mathcal N_u$
shown in the proof of Corollary~\ref{thm:tools-bult},
$\mathcal N_u(\boldsymbol h';p,q)=\mathcal N_u(\boldsymbol h;p,q)$, so
\begin{equation}
 \frac{E(u;\boldsymbol h;p,q)}{\mathcal N_u(\boldsymbol h;p,q)}
 =\frac{E(u;\boldsymbol h';p,q)}{\mathcal N_u(\boldsymbol h';p,q)},
 \label{eq:ad-bult-normalization}
\end{equation}
and the two integrals have the same measure $\kappa/2$ and no
relative factor.

\subsection{Reflection cancellations and the index identity}

By \eqref{eq:tools-reflection}, each boundary quotient in
\eqref{eq:tools-bult-E} becomes
\begin{equation}
 \frac{\Gamma(h_rz^{\pm1})}{\Gamma(uh_rz^{\pm1})}
 =\Gamma\bigl(h_rz^{\pm1},Pz^{\pm1}/(uh_r)\bigr).
 \label{eq:ad-paired-reflection}
\end{equation}
After removing the common factor $P^{1/2}$, the eight boundary
parameters for $\boldsymbol h$ are
\begin{equation}
 \bigl(u^{-2}x,ux^{-1},u^{-1}x,x^{-1},x,u^{-1}x^{-1},
       u^{-2}x^{-1},ux\bigr),
 \label{eq:ad-h1-original-eight}
\end{equation}
and those for $\boldsymbol h'$ are
\begin{equation}
 \bigl(u^{-1/2},u^{-1/2},u^{1/2},u^{-3/2},u^{3/2},u^{-5/2},
       u^{-1/2}x^2,u^{-1/2}x^{-2}\bigr).
 \label{eq:ad-h1-transformed-eight}
\end{equation}
Reflection cancels two entries whenever their residual factors
multiply to one. In \eqref{eq:ad-h1-original-eight}, the pairs
\begin{equation}
 (ux^{-1},u^{-1}x),\qquad (x^{-1},x),\qquad
 (u^{-1}x^{-1},ux)
\end{equation}
cancel, leaving $u^{-2}x$ and $u^{-2}x^{-1}$. In
\eqref{eq:ad-h1-transformed-eight}, one occurrence of $u^{-1/2}$
cancels against $u^{1/2}$, and $u^{-3/2}$ cancels against $u^{3/2}$.
The remaining entries are
\begin{equation}
 u^{-5/2},\qquad u^{-1/2},\qquad
 u^{-1/2}x^2,\qquad u^{-1/2}x^{-2}.
\end{equation}
The adjoint and vector factors are unchanged. Comparing with
\eqref{eq:ad-h1-two}--\eqref{eq:ad-h1-four} gives
\begin{equation}
 \frac{E(u;\boldsymbol h;p,q)}{\mathcal N_u(\boldsymbol h;p,q)}
 =F_2(x),\qquad
 \frac{E(u;\boldsymbol h';p,q)}{\mathcal N_u(\boldsymbol h';p,q)}
 =F_4(x^2).
\end{equation}
Equation~\eqref{eq:ad-bult-normalization} then proves
\eqref{eq:ad-h1-identity}, and multiplication by the common factor
$\Gamma(u^4)\Gamma(u)/\Gamma(u^2)$ in \eqref{eq:ad-h1-prefactor} gives
the equality of the interacting indices, with the free $M_1$ removed
and the flavor map $y=x^2$.

The unit circle is an admissible contour for $|x|=1$ and
\begin{equation}
 |P|^{1/5}<|u|<1,
 \label{eq:ad-h1-contour-domain}
\end{equation}
since the boundary parameter of largest modulus is
$P^{1/2}u^{-5/2}$, so all eight parameters on either side, and the
adjoint parameter, have modulus less than one. This persists in an
open neighborhood, and meromorphic continuation gives the general
identity.

Since multiplication by common spectator factors and integration over
a common flavor subgroup preserve the equality on a shared regular
contour domain, an $H_1$ block can be replaced inside a larger quiver. This
does not give the full-index identities for the $n\ge2$ Cho--Song
quivers.

\subsection{The generalized Coulomb limit and the Macdonald constant term}
\label{subsec:ad-two-parameter}

The generalized Coulomb limit retains two parameters:
\begin{equation}
 \xi=P/t,\qquad
 \mathcal J(p,\xi;\boldsymbol f)
 =\lim_{q\to0}\mathcal I(p,q,t=pq/\xi;\boldsymbol f),
 \qquad 0<|p|<1,\quad 0<|\xi|<1,
 \label{eq:ad-two-parameter-limit}
\end{equation}
where $\boldsymbol f$ denotes the flavor fugacities and fractional
powers are chosen consistently. Gauge-theory indices in this limit
were studied in \cite[Section~8]{Gadde:2011uv}. The variables $(q,u)$ of
\cite[Eq.~(3.23) in the published version]{Buican:2023efi} are our
$(p,\xi)$ after exchanging the two nomes, and $p=0$ gives the
ordinary Coulomb index. This limit differs from the Schur limit $t=q$
and from the Macdonald limit $p\to0$ at fixed $q,t$; for the latter,
Kang, Lawrie and Song proposed an algebraic construction of the
Macdonald indices of Argyres--Douglas theories
\cite{Kang:2026ona,Kang:2026ney}.

\paragraph{The complete starting integrals.}
We use the four family labels
\[
 \begin{aligned}
 D_{\mathrm o}&=(A_1,D_{2n+1}),&
 A_{\mathrm e}&=(A_1,A_{2n}),\\
 D_{\mathrm e}&=(A_1,D_{2n+2}),&
 A_{\mathrm o}&=(A_1,A_{2n+1}).
 \end{aligned}
\]
The label $\mathfrak f$ denotes any one of these families and
$\mathrm Q$ its $SU(2)^n$ quiver. Both descriptions are written below,
following \cite[Sections~2.1--3.2]{Cho:2026tqf}. Throughout this
subsection,
$\Gamma(a)=\Gamma(a;p,q)$, $\Gamma(a_1,\ldots,a_s)=\prod_j\Gamma(a_j)$,
and $Sp(n)$ in that reference is written as $USp(2n)$.
Independent signs have the convention of \eqref{eq:tools-G};
parentheses instead specify a correlated pair,
\[
 \Gamma\bigl(h(yb)^{\pm1}\bigr)=\Gamma(hyb,h/(yb)).
\]
The unit contours below are admissible for sufficiently small $q$
along \eqref{eq:ad-two-parameter-limit}, as shown below.

\paragraph{The four quiver integrals.}
With $\kappa=\kappa(p,q)$ as in \eqref{eq:tools-nomes}, define the
local $SU(2)$ factor
\begin{equation}\label{eq:ad-start-localblock}
\mathcal B(w;z)=\frac{\kappa}{2}
\frac{\Gamma(wz^2,w,wz^{-2})}{\Gamma(z^2,z^{-2})}.
\end{equation}
For adjoint weights $\xi^{\alpha_i}$, all four quiver indices take the
form
\begin{equation}\label{eq:ad-start-fullQ}
\boxed{\mathcal I_{\mathfrak f,\mathrm Q}
=S_{\mathfrak f,\mathrm Q}(\xi)\oint_{\mathbb T^n}
\prod_{i=1}^n\left[\frac{\dd z_i}{2\pi\mathrm i z_i}\mathcal B(\xi^{\alpha_i};z_i)\right]
L_{\mathfrak f,\mathrm Q}(z;t,\xi;\boldsymbol f).}
\end{equation}
In $\mathcal B$, the factor $\Gamma(w)$ is the adjoint zero weight and
$1/2$ is the Weyl normalization. The prefactor $S$ contains the
singlets and reflected Casimir flips, and $L$ the fundamental and
bifundamental factors.

For $D_{\mathrm o}$, put $d=2n+1$ and $z_0=a$, where $a$ is the left
flavor fugacity. Then
\begin{align}
\alpha_i&=\frac{d-2i}{d},\quad 1\le i\le n,\\
S_{D_{\mathrm o},\mathrm Q}&=
\frac{\displaystyle\prod_{s=1}^{\lceil n/2\rceil}
\Gamma\!\left(\xi^{2-2(2s-1)/d}\right)}
{\displaystyle\prod_{k=\lfloor n/2\rfloor+1}^{n}
\Gamma\!\left(\xi^{2(d-2k)/d}\right)},\label{eq:ad-start-SDo}\\
L_{D_{\mathrm o},\mathrm Q}&=\prod_{i=1}^n
\Gamma\!\left(t^{1/2}\xi^{(4i-2n-3)/(2d)}
z_{i-1}^{\pm1}z_i^{\pm1}\right).\label{eq:ad-start-LDo}
\end{align}
The $i=1$ factor is a pair of gauge fundamentals with flavor weights
$a^{\pm1}$, and each $i\ge2$ gives a bifundamental between neighboring
nodes.

For $A_{\mathrm e}$, put $d=2n+3$. The two left fundamentals are
split, with no flavor parameter in this presentation:
\begin{align}
\alpha_i&=\frac{2n+1-2i}{d},\quad1\le i\le n,\\
S_{A_{\mathrm e},\mathrm Q}&=
\frac{\displaystyle\Gamma\!\left(\xi^{2(2n+1)/d}\right)
\prod_{s=1}^{\lfloor n/2\rfloor}
\Gamma\!\left(\xi^{4(n+1-s)/d}\right)}
{\displaystyle\prod_{k=\lceil n/2\rceil}^{n}
\Gamma\!\left(\xi^{2(2n+1-2k)/d}\right)},\label{eq:ad-start-SAe}\\
L_{A_{\mathrm e},\mathrm Q}&=
\Gamma\!\left(t^{1/2}\xi^{(1-2n)/d}z_1^{\pm1}\right)
\Gamma\!\left(t^{1/2}\xi^{2/d}z_1^{\pm1}\right)\notag\\
&\quad\times\prod_{i=2}^{n}
\Gamma\!\left(t^{1/2}\xi^{(4i-2n-1)/(2d)}
z_{i-1}^{\pm1}z_i^{\pm1}\right).\label{eq:ad-start-LAe}
\end{align}
For $n\ge2$ the $s=1$ term in the singlet product is $M_1$. For $n=1$
this product is empty, so the free $M_1$ is absent and only $M_0$
remains in the numerator.

For $D_{\mathrm e}$, put $d=n+1$ and $z_0=a$. There is a flavor doublet
at each end, with right fugacity $x$:
\begin{align}
\alpha_i&=\frac{n+1-i}{d},\quad1\le i\le n,\\
S_{D_{\mathrm e},\mathrm Q}&=
\frac{\displaystyle\prod_{s=1}^{\lceil n/2\rceil}
\Gamma\!\left(\xi^{(2n+3-2s)/d}\right)}
{\displaystyle\prod_{k=\lfloor n/2\rfloor+1}^{n}
\Gamma\!\left(\xi^{2(n+1-k)/d}\right)},\label{eq:ad-start-SDe}\\
L_{D_{\mathrm e},\mathrm Q}&=\left[\prod_{i=1}^n
\Gamma\!\left(t^{1/2}\xi^{(2i-n-2)/(2d)}
z_{i-1}^{\pm1}z_i^{\pm1}\right)\right]
\Gamma\!\left(t^{1/2}\xi^{n/(2d)}z_n^{\pm1}x^{\pm1}\right).
\label{eq:ad-start-LDe}
\end{align}

For $A_{\mathrm o}$, put $d=n+2$. The left pair is split, and a flavor
doublet remains on the right:
\begin{align}
\alpha_i&=\frac{n+1-i}{d},\quad1\le i\le n,\\
S_{A_{\mathrm o},\mathrm Q}&=
\frac{\displaystyle\Gamma\!\left(\xi^{(2n+2)/d}\right)
\prod_{s=1}^{\lfloor n/2\rfloor}
\Gamma\!\left(\xi^{(2n+3-2s)/d}\right)}
{\displaystyle\prod_{k=\lceil n/2\rceil}^{n}
\Gamma\!\left(\xi^{2(n+1-k)/d}\right)},\label{eq:ad-start-SAo}\\
L_{A_{\mathrm o},\mathrm Q}&=
\Gamma\!\left(t^{1/2}\xi^{-n/d}z_1^{\pm1}\right)
\Gamma\!\left(t^{1/2}\xi^{1/d}z_1^{\pm1}\right)\notag\\
&\quad\times\prod_{i=2}^n
\Gamma\!\left(t^{1/2}\xi^{(2i-n-1)/(2d)}
z_{i-1}^{\pm1}z_i^{\pm1}\right)
\Gamma\!\left(t^{1/2}\xi^{(n+1)/(2d)}z_n^{\pm1}x^{\pm1}\right).
\label{eq:ad-start-LAo}
\end{align}
As for $A_{\mathrm e}$, at $n=1$ the singlet product over $s\ge1$ is
empty, so the free $M_1$ is absent. Products over empty ranges equal
one. For $n=1$, both end factors in \eqref{eq:ad-start-LDe} or
\eqref{eq:ad-start-LAo} couple to the same node and must both be
retained.

\paragraph{The single-group adjoint integrals.}
For $G=USp(2n)$ define
\begin{align}
\mathfrak A_{C_n}(u;L)
&=\frac{\kappa^n\Gamma(u)^n}{2^n n!}
\oint_{\mathbb T^n}\prod_{i=1}^n\frac{\dd y_i}{2\pi iy_i}
\prod_{i=1}^n\frac{\Gamma(uy_i^{\pm2})}{\Gamma(y_i^{\pm2})}
\notag\\[-2pt]
&\hspace{12mm}\times
\prod_{1\le i<j\le n}
\frac{\Gamma(uy_i^{\pm1}y_j^{\pm1})}
{\Gamma(y_i^{\pm1}y_j^{\pm1})}\,L(y).
\label{eq:ad-start-AC}
\end{align}
All roots, long and short, carry the same nomes $(p,q)$; $\Gamma(u)^n$
is the contribution of the $n$ adjoint zero weights, and $2^nn!$ is
the order of the Weyl group.

For $G=SU(n+1)$ define $y_{n+1}=(y_1\cdots y_n)^{-1}$ and
\begin{equation}\label{eq:ad-start-AA}
\mathfrak A_{A_n}(u;L)
=\frac{\kappa^n\Gamma(u)^n}{(n+1)!}
\oint_{\mathbb T^n}\prod_{i=1}^n\frac{\dd y_i}{2\pi iy_i}
\prod_{\substack{1\le i,j\le n+1\\i\ne j}}
\frac{\Gamma(uy_i/y_j)}{\Gamma(y_i/y_j)}\,L(y).
\end{equation}
The product runs over ordered pairs, that is, over all roots.

\paragraph{The single-group integrals.}
The families $D_{\mathrm o},A_{\mathrm e}$ have $G=USp(2n)$ and
$R_G=C_n$, whereas $D_{\mathrm e},A_{\mathrm o}$ have $G=SU(n+1)$
and $R_G=A_n$. With $u=\xi^{1/d}$ and $d$ as above, the indices are
\begin{align}
\mathcal I_{\mathfrak f,G}
&=S_{\mathfrak f,G}(\xi)\,
\mathfrak A_{R_G}(u;L_{\mathfrak f,G}),\label{eq:ad-start-fullG}\\
S_{\mathfrak f,G}(\xi)
&=\frac{\prod_{j=1}^n\Gamma(\xi^{\Delta_j})}
{\prod_{j=1}^n\Gamma(u^{e_j+1})},\label{eq:ad-start-SG}
\end{align}
where $e_j=2j-1$ for $C_n$ and $e_j=j$ for $A_n$, and $\Delta_j$ are
the dimensions in \eqref{eq:ad-coulomb-spectra} below. Explicitly,
\begin{align}
S_{D_{\mathrm o},G}&=\frac{\prod_{m=1}^n\Gamma(\xi^{2-2m/(2n+1)})}
{\prod_{j=1}^n\Gamma(\xi^{2j/(2n+1)})},\\
S_{A_{\mathrm e},G}&=\frac{\prod_{m=n+2}^{2n+1}\Gamma(\xi^{2m/(2n+3)})}
{\prod_{j=1}^n\Gamma(\xi^{2j/(2n+3)})},\\
S_{D_{\mathrm e},G}&=\frac{\prod_{m=n+2}^{2n+1}\Gamma(\xi^{m/(n+1)})}
{\prod_{j=2}^{n+1}\Gamma(\xi^{j/(n+1)})},\\
S_{A_{\mathrm o},G}&=\frac{\prod_{m=n+3}^{2n+2}\Gamma(\xi^{m/(n+2)})}
{\prod_{j=2}^{n+1}\Gamma(\xi^{j/(n+2)})}.
\end{align}
The corresponding matter factors are
\begin{align}
L_{D_{\mathrm o},G}
&=\prod_{i=1}^n
\Gamma\!\left(t^{1/2}\xi^{n/d}y_i^{\pm1}a^{0,\pm2}\right)
\Gamma\!\left(t^{1/2}\xi^{-n/d}y_i^{\pm1}\right),
\quad d=2n+1,\label{eq:ad-start-LGDo}\\
L_{A_{\mathrm e},G}
&=\prod_{i=1}^n
\Gamma\!\left(t^{1/2}\xi^{(n+1)/d}y_i^{\pm1}\right)
\Gamma\!\left(t^{1/2}\xi^{-n/d}y_i^{\pm1}\right),
\quad d=2n+3,\label{eq:ad-start-LGAe}\\
L_{D_{\mathrm e},G}
&=\prod_{i=1}^{n+1}
\Gamma\!\left(t^{1/2}\xi^{n/(2d)}
(y_i b_1 b_2^{2n+1})^{\pm1}\right)\notag\\
&\hspace{12mm}\times
\Gamma\!\left(t^{1/2}\xi^{-n/(2d)}
(y_i b_1 b_2^{-1})^{\pm1}\right),
\quad d=n+1,\label{eq:ad-start-LGDe}\\
L_{A_{\mathrm o},G}
&=\prod_{i=1}^{n+1}
\Gamma\!\left(t^{1/2}\xi^{-n/(2d)}(y_i b_3)^{\pm1}\right),
\quad d=n+2.\label{eq:ad-start-LGAo}
\end{align}
In \eqref{eq:ad-start-LGDo}, $a^{0,\pm2}$ denotes the three flavor
monomials $1,a^2,a^{-2}$, combined independently with $y_i^{\pm1}$. In
\eqref{eq:ad-start-LGDe}--\eqref{eq:ad-start-LGAo} the paired signs
give a fundamental and an antifundamental. The flavor parameters
$b_1,b_2,b_3$ are named differently from those of the quivers; their
identification is not needed below, since the limits are independent
of all flavor parameters.

\paragraph{Matter degeneration and the common contours.}
As $q\to0$ with $v,c$ fixed, nonzero and away from poles,
\eqref{eq:tools-gamma} and reflection give
\begin{equation}
 \begin{aligned}
 \Gamma(v;p,q)&\longrightarrow\frac1{(v;p)_\infty},
 &\Gamma(P^{1/2}c;p,q)&\longrightarrow1,\\
 \Gamma(P/v;p,q)&=\frac1{\Gamma(v;p,q)}
 \longrightarrow(v;p)_\infty.
 \end{aligned}
 \label{eq:ad-two-parameter-gamma-limit}
\end{equation}
In the first limit only the $q^0$ layer of the denominator survives,
and in the second both the argument and its reflection partner tend
to zero. The third limit applies to the Casimir flips, whose operator
weight $v$ is fixed, and the first to the singlets $M_s$, which have
fixed weights $\xi^{\mu_s}$ even when they flip mesons.

Every fundamental or bifundamental argument in
\eqref{eq:ad-start-LDo}--\eqref{eq:ad-start-LAo} and
\eqref{eq:ad-start-LGDo}--\eqref{eq:ad-start-LGAo} has the form
\begin{equation}
 t^{1/2}\xi^b z^\nu\boldsymbol f^\omega
 =P^{1/2}\underbrace{\xi^{b-1/2}z^\nu
             \boldsymbol f^\omega}_{c(z,\boldsymbol f)},
 \qquad t=P/\xi.
 \label{eq:ad-matter-substitution}
\end{equation}
Here $z^\nu$ denotes the appropriate monomial in the quiver or
single-group gauge variables, and $\boldsymbol f^\omega$ a flavor
monomial. For example, a $D_{\mathrm o}$ edge gives
\[
 t^{1/2}\xi^{(4i-2n-3)/(2d)}
 =P^{1/2}\xi^{-2(n+1-i)/d},\qquad d=2n+1.
\]
The power of $\xi$ may be negative, but at fixed $\xi\ne0$ it is
finite and independent of $q$.

Fix $n$, $0<|p|,|\xi|<1$, and flavor fugacities in a compact set
$K\subset(\mathbb C^*)^s$. On the unit gauge tori $c$ and $c^{-1}$
are uniformly bounded, so $\Gamma(P^{1/2}c;p,q)=1+O(|q|^{1/2})$
uniformly and
\begin{equation}
 L_{\mathfrak f,\mathrm Q}\longrightarrow1,\qquad
 L_{\mathfrak f,G}\longrightarrow1
 \qquad(q\to0,\ t=pq/\xi).
 \label{eq:ad-matter-to-one}
\end{equation}
The flavor dependence and the couplings between nodes drop
out. For small $q$ the unit tori are common contours, since the fixed
adjoint coefficients have modulus less than one and the matter
coefficients tend to zero. The remaining product limits are uniform
there: adjoint denominators stay away from zero, vector reciprocals
are regular at Weyl walls, and scalar prefactors remain bounded. The
integrand $F_q$ thus converges uniformly, and
\begin{equation}
 \lim_{q\to0}\int_{\mathbb T^n}F_q\,\dd\mu
 =\int_{\mathbb T^n}\lim_{q\to0}F_q\,\dd\mu,
 \qquad \dd\mu=\prod_{\ell=1}^n\frac{\dd\theta_\ell}{2\pi},
 \label{eq:ad-limit-under-integral}
\end{equation}
where $z_\ell=e^{\mathrm i\theta_\ell}$.

\paragraph{The normalized constant term.}
For a compact connected simple group $G$ of rank $r$, let $R$ be
its root system, $W$ its Weyl group, and $e_1,\ldots,e_r$ its
exponents. Write $\operatorname{CT}$ for the constant Laurent
coefficient, equivalently integration over the unit gauge tori
with measure $\prod_{\ell=1}^r\dd z_\ell/(2\pi\mathrm i z_\ell)$.

\begin{lemma}[Normalized Macdonald constant term]
\label{lem:ad-macdonald-ct}
For $|p|,|w|<1$,
\begin{align}
 \mathcal M_G(p,w)
 &:=\frac{(p;p)_\infty^r}{|W|(w;p)_\infty^r}
 \operatorname{CT}\prod_{\alpha\in R}
 \frac{(z^\alpha;p)_\infty}{(wz^\alpha;p)_\infty}
 \notag\\
 &=\prod_{j=1}^{r}
 \frac{(p w^{e_j};p)_\infty}{(w^{e_j+1};p)_\infty}.
 \label{eq:ad-normalized-macdonald-ct}
\end{align}
\end{lemma}

This is the Macdonald constant-term identity, proved by Cherednik
\cite{MR1224114}. For the normalization, consider
Stembridge's kernel \cite{MR2199453}
\begin{equation}
 K_G(p,w;z)=
 \left[\frac{(p;p)_\infty}{(w;p)_\infty}\right]^r
 \prod_{\alpha\in R}
 \frac{(pz^\alpha;p)_\infty}{(wz^\alpha;p)_\infty}.
\end{equation}
Its trivial-character coefficient $C_0(p,w)$ is the product on
the right-hand side of \eqref{eq:ad-normalized-macdonald-ct}, by
\cite[Eq.~(1.5)]{MR2199453}. Normalized Haar averaging
and the Weyl integration formula give
\begin{align}
 C_0(p,w)
 &=\frac1{|W|}\operatorname{CT}
   \left[\prod_{\alpha\in R}(1-z^\alpha)K_G(p,w;z)\right]
   \notag\\
 &=\frac{(p;p)_\infty^r}{|W|(w;p)_\infty^r}
   \operatorname{CT}\prod_{\alpha\in R}
   \frac{(z^\alpha;p)_\infty}{(wz^\alpha;p)_\infty}
   =\mathcal M_G(p,w),
 \label{eq:ad-macdonald-haar}
\end{align}
using $(1-a)(pa;p)_\infty=(a;p)_\infty$. This is the gauge-index
normalization of \cite[Eqs.~(8.4)--(8.8)]{Gadde:2011uv}, with the same base
$p$ for the long and short roots of $C_n$.

This normalization also follows directly from the starting
integrals: $\kappa^r\to(p;p)_\infty^r$, and the $r$ adjoint zero
weights give $\Gamma(w)^r\to(w;p)_\infty^{-r}$. In particular,
\begin{equation}
 SU(n+1):\ \{e_j\}=\{1,2,\ldots,n\},\qquad
 USp(2n):\ \{e_j\}=\{1,3,\ldots,2n-1\}.
 \label{eq:ad-group-exponents}
\end{equation}
The basic adjoint invariants have degrees $e_j+1$, namely
$2,3,\ldots,n+1$ and $2,4,\ldots,2n$, respectively.
An $SU(2)$ node with adjoint weight $\xi^\alpha$ therefore contributes
\begin{equation}
 \mathcal M_{SU(2)}(p,\xi^\alpha)
 =\frac{(p\xi^\alpha;p)_\infty}{(\xi^{2\alpha};p)_\infty}.
 \label{eq:ad-node-limit}
\end{equation}
At $p=0$, the lemma reduces to the adjoint Molien average
$\prod_j(1-w^{e_j+1})^{-1}$.

\subsection{The four arbitrary-rank identities}
\label{subsec:ad-all-rank}

In this limit the rank-one $H_0$ index follows from the analysis of
Bhargava, Buican and Jiang \cite{Bhargava:2022yik}. For a freely
generated Coulomb branch, Buican and Jiang conjectured a factor in the
generalized Coulomb index for each generator of dimension $\Delta$
\cite[Conjecture~4, Eq.~(3.32) in the published version]{Buican:2023efi},
which in our variables is
\[
 \operatorname{PE}\!\left[\frac{(\xi-p)\xi^{\Delta-1}}{1-p}\right]
 =\frac{(p\xi^{\Delta-1};p)_\infty}{(\xi^\Delta;p)_\infty},
\]
and discussed the rank-one examples $H_0,H_1,H_2$
\cite[Eqs.~(3.28)--(3.30)]{Buican:2023efi}. We evaluate both ultraviolet
integrals for each family, with the Coulomb spectra of
\cite[Sections~2.1--3.2]{Cho:2026tqf}.

\begin{theorem}[Two-parameter identities for all four families]
\label{thm:ad-two-parameter}
For every positive integer $n$, the $SU(2)^n$ quiver and single-group
descriptions of each family $\mathfrak f$ have the same limit
\eqref{eq:ad-two-parameter-limit}:
\begin{equation}
 \boxed{\displaystyle
 \mathcal J_{\mathfrak f,\mathrm Q}(p,\xi)
 =\mathcal J_{\mathfrak f,G}(p,\xi)
 =\prod_{j=1}^{n}
 \frac{(p\xi^{\Delta_j-1};p)_\infty}
      {(\xi^{\Delta_j};p)_\infty}.}
 \label{eq:ad-two-parameter-result}
\end{equation}
The ordinary flavor fugacities drop out. The dimensions are
\begin{equation}
 \begin{array}{c|c}
 \text{family}&\{\Delta_1,\ldots,\Delta_n\}\\ \hline
 (A_1,D_{2n+1})&
 \{2-2m/(2n+1):1\le m\le n\}\\
 (A_1,A_{2n})&
 \{2m/(2n+3):n+2\le m\le2n+1\}\\
 (A_1,D_{2n+2})&
 \{m/(n+1):n+2\le m\le2n+1\}\\
 (A_1,A_{2n+1})&
 \{m/(n+2):n+3\le m\le2n+2\}.
 \end{array}
 \label{eq:ad-coulomb-spectra}
\end{equation}
The interacting-index prescription is understood: at $n=1$ the
$M_1$ singlet in the even-$A$ and odd-$A$ families is removed as
in \cite{Cho:2026tqf}. The identity holds initially in the domain
of \eqref{eq:ad-two-parameter-limit} and extends meromorphically
on the chosen branches, equivalently in $u=\xi^{1/d}$ with the
family-dependent $d$, wherever both sides are defined.
\end{theorem}

\begin{proof}
Use the adjoint exponents $\alpha_i$ and singlet prefactors in
\eqref{eq:ad-start-SDo}--\eqref{eq:ad-start-SAo}.
Let $\mathcal F$ be the set of nodes with a flipped adjoint Casimir,
and let $\mathcal S$ label the surviving singlets of weights
$\xi^{\mu_s}$. Write $S^{(0)}$ for the limit of a singlet
prefactor at fixed $p,\xi$. By \eqref{eq:ad-matter-to-one}, the
matter factor in \eqref{eq:ad-start-fullQ} tends to one, so the
quiver integral factorizes. Equation~\eqref{eq:ad-node-limit} gives
\begin{align}
 \mathcal J_{\mathfrak f,\mathrm Q}
 &=S^{(0)}_{\mathfrak f,\mathrm Q}
   \prod_{i=1}^n\mathcal M_{SU(2)}(p,\xi^{\alpha_i})\notag\\
 &=\frac{\prod_{k\in\mathcal F}(\xi^{2\alpha_k};p)_\infty}
         {\prod_{s\in\mathcal S}(\xi^{\mu_s};p)_\infty}
   \prod_{i=1}^n
   \frac{(p\xi^{\alpha_i};p)_\infty}
        {(\xi^{2\alpha_i};p)_\infty}\notag\\
 &=\frac{\prod_{i=1}^{n}(p\xi^{\alpha_i};p)_\infty}
 {\displaystyle
  \prod_{i\notin\mathcal F}(\xi^{2\alpha_i};p)_\infty
  \prod_{s\in\mathcal S}(\xi^{\mu_s};p)_\infty}.
 \label{eq:ad-quiver-limit-product}
\end{align}
Each Casimir flip removes its node's denominator, and the surviving
$M_s$ supply the remaining singlet denominators.

For the single-group integral \eqref{eq:ad-start-fullG}, put
$u=\xi^{1/d}$. The prefactor \eqref{eq:ad-start-SG} flips every
basic adjoint Casimir. The same matter limit and
Lemma~\ref{lem:ad-macdonald-ct} therefore give
\begin{align}
 \mathcal J_{\mathfrak f,G}
 &=S^{(0)}_{\mathfrak f,G}\,\mathcal M_G(p,u)\notag\\
 &=\frac{\prod_{j=1}^n(u^{e_j+1};p)_\infty}
         {\prod_{j=1}^n(\xi^{\Delta_j};p)_\infty}
   \prod_{j=1}^n
   \frac{(pu^{e_j};p)_\infty}{(u^{e_j+1};p)_\infty}\notag\\
 &=\frac{\prod_{j=1}^{n}(p u^{e_j};p)_\infty}
       {\prod_{j=1}^{n}(\xi^{\Delta_j};p)_\infty}.
 \label{eq:ad-single-group-limit-product}
\end{align}
It remains to match the multisets of exponents.

For the quiver denominator, we verify
\begin{equation}
 \{2\alpha_i:i\notin\mathcal F\}
 \sqcup\{\mu_s:s\in\mathcal S\}
 =\{\Delta_j:1\le j\le n\}.
\end{equation}
Here $\sqcup$ denotes multiset union, and empty index ranges
contribute no entries. The four prefactors give the following
integer-label partitions.

For $D_{\mathrm o}$, write each exponent as $2-2m/(2n+1)$.
The unflipped Casimirs and singlets, respectively, give
\begin{equation}
 \{2i:1\le i\le\lfloor n/2\rfloor\}
 \sqcup\{2s-1:1\le s\le\lceil n/2\rceil\}
 =\{1,\ldots,n\}.
\end{equation}
For $A_{\mathrm e}$, write the exponents as $2m/(2n+3)$.
The Casimirs, the $M_s$ with $s\ge1$, and $M_0$ give
\begin{equation}
 \begin{aligned}
 &\{2n+1-2i:1\le i<\lceil n/2\rceil\}\\
 &\quad\sqcup\{2(n+1-s):1\le s\le\lfloor n/2\rfloor\}
 \sqcup\{2n+1\}
 =\{n+2,\ldots,2n+1\}.
 \end{aligned}
\end{equation}
For $D_{\mathrm e}$, write the exponents as $m/(n+1)$.
The Casimirs and singlets give
\begin{equation}
 \begin{aligned}
 &\{2(n+1-i):1\le i\le\lfloor n/2\rfloor\}\\
 &\quad\sqcup\{2n+3-2s:1\le s\le\lceil n/2\rceil\}
 =\{n+2,\ldots,2n+1\}.
 \end{aligned}
\end{equation}
For $A_{\mathrm o}$, write the exponents as $m/(n+2)$.
The Casimirs, the $M_s$ with $s\ge1$, and $M_0$ give
\begin{equation}
 \begin{aligned}
 &\{2(n+1-i):1\le i<\lceil n/2\rceil\}\\
 &\quad\sqcup\{2n+3-2s:1\le s\le\lfloor n/2\rfloor\}
 \sqcup\{2n+2\}
 =\{n+3,\ldots,2n+2\}.
 \end{aligned}
\end{equation}
In each line the two progressions occupy opposite parity classes;
in the $A$ families, $M_0$ supplies the last missing endpoint.
The lists are disjoint and exhaust the indicated interval, so
every denominator exponent occurs once. At $n=1$, only $M_0$
remains in each $A$ family: the free $M_1$ has already been
removed from \eqref{eq:ad-start-SAe} and \eqref{eq:ad-start-SAo}.
The odd-$A$ removal agrees with \eqref{eq:ad-h1-prefactor}.

Finally the numerator multisets agree, since
\begin{equation}
 \begin{array}{c|c|c|c}
 \text{family}&G&d&
 \{d\alpha_i\}=\{e_i\}=\{d(\Delta_j-1)\}\\ \hline
 (A_1,D_{2n+1})&USp(2n)&2n+1&\{1,3,\ldots,2n-1\}\\
 (A_1,A_{2n})&USp(2n)&2n+3&\{1,3,\ldots,2n-1\}\\
 (A_1,D_{2n+2})&SU(n+1)&n+1&\{1,2,\ldots,n\}\\
 (A_1,A_{2n+1})&SU(n+1)&n+2&\{1,2,\ldots,n\}.
 \end{array}
 \label{eq:ad-exponent-matching}
\end{equation}
Equations \eqref{eq:ad-quiver-limit-product} and
\eqref{eq:ad-single-group-limit-product} therefore both equal
\eqref{eq:ad-two-parameter-result}.
\end{proof}

\begin{corollary}[Coulomb specialization]
\label{cor:ad-coulomb}
Both descriptions have the same Coulomb index at every positive rank:
\begin{equation}
 \mathcal I_C(\xi)=\mathcal J(0,\xi)
 =\prod_{j=1}^{n}\frac1{1-\xi^{\Delta_j}},
 \label{eq:ad-coulomb-result}
\end{equation}
with the four spectra given in \eqref{eq:ad-coulomb-spectra}.
\end{corollary}

This follows from $(v;0)_\infty=1-v$, since the numerators in
\eqref{eq:ad-two-parameter-result} tend to one as $p\to0$. For
example, for the rank-two $(A_1,D_5)$ theory,
\begin{equation}
 \mathcal J_{(A_1,D_5)}(p,\xi)=
 \frac{(p\xi^{1/5};p)_\infty(p\xi^{3/5};p)_\infty}
      {(\xi^{6/5};p)_\infty(\xi^{8/5};p)_\infty}.
 \label{eq:ad-d5-two-parameter}
\end{equation}
At nonzero $q$ the bifundamentals couple the gauge nodes and the
constant-term evaluation no longer applies, so the full and
Schur-index identities for $n\geq2$ remain open.

\section{Conclusion}\label{sec:conclusion}

We have derived index identities for three classes of four-dimensional
dualities by specializing established integral transformations. For
$B_n/C_n$, the flavored Schur indices agree at every rank: theta
function duplication identifies the two root systems inside Rains's
diagonal trace transformation, and a regulated contour limit completes
the comparison. For Kutasov--Schwimmer duality, the identity holds at
independent nomes for $F=2$, $N=k+1$, $M=k-1$ in the ultraviolet
grading of Section~\ref{sec:ks}, and on $p=q^k$ for balanced flavor
families that admit a nonempty open set of regular parameters. For the
Argyres--Douglas quivers, van de Bult's transformation proves the full
flavored rank-one $H_1$ identity, and at arbitrary rank both
descriptions of all four families give the generalized Coulomb index
\eqref{eq:ad-two-parameter-result} conjectured by Buican and Jiang
\cite{Buican:2023efi}, which reduces at $p=0$ to the Coulomb generating
function.

Each of these results leaves an identity at generic fugacities open.

\begin{enumerate}
\item \emph{Generic $B_n/C_n$ indices.}
Spiridonov and Vartanov formulated the full elliptic comparison with
independent $p,q$ and three adjoint weights $u,v,w$, $uvw=P$, and
studied its degenerations, including the Macdonald limit \cite{Spiridonov:2010qv}.
The next case is the deformed Schur, or flavor-refined Macdonald,
identity of \cite{Ren:2025tvx}, with root factor
\begin{equation}
 \frac{(z;q)_\infty(uvz;q)_\infty}
 {(uz;q)_\infty(vz;q)_\infty}
 \label{eq:conclusion-macdonald-weight}
\end{equation}
and $q,u,v$ independent. Our proof covers $q=s^2$, $uv=q$, and the
Macdonald constant-term evaluation covers $v=0$; a common system of
difference equations with a uniqueness property might connect these
loci to the generic case.

\item \emph{Kutasov--Schwimmer duality at finite rank.}
The general identity relates $N$ and $M=kF-N$ at independent $p,q$
with all flavor fugacities \cite{Kutasov:1995np,Spiridonov:2009za}. A simple
interacting example outside our independent-nome $F=2$ family is
$k=2$, $N=3$, $F=4$, $M=5$, with $R(X)=2/3$ and $R(Q)=1/2$, for which
all elementary gauge-invariant chiral operators lie above the
free-field bound. The deconfinement results of \cite{Benvenuti:2024glr} establish
related three-dimensional adjoint and four-dimensional symplectic
antisymmetric cases, not this general unitary identity; extending the
sequence of elementary transformations used here is one way to
approach it.

\item \emph{Argyres--Douglas quivers at higher rank.}
For the rank-two $(A_1,D_5)$ theory, Cho and Song compare the
$SU(2)\times SU(2)$ and $USp(4)$ descriptions by finite series
expansions and ask for an analytic proof \cite{Cho:2026tqf};
Theorem~\ref{thm:ad-two-parameter} proves its two-parameter
degeneration. At the Schur point, Jiang found identities relating
different Argyres--Douglas theories after a change of flavor and
grading variables \cite{Jiang:2026ild}; here the question is whether the
two descriptions of the same theory agree at generic fugacities.
\end{enumerate}
In all three classes of theories, although we are not able to prove index identities in full generality, we perform some 
additional tests beyond the proven cases with low order expansions and find no inconsistency.  

The same methods may apply to other dual pairs. Duan and Zhong
catalogued pairs of class~$\mathcal S$ theories of untwisted type~$A$
and twisted type~$D$, and checked their Schur and Hall--Littlewood
indices to finite order \cite{Duan:2026hhj}. Hatsuda, Lin and Okazaki
matched half-indices for dual boundaries and interfaces
\cite{Hatsuda:2024lcc} and computed line-defect correlators
for orthogonal and symplectic gauge groups
\cite{Hatsuda:2025jze}; it would be interesting to apply the
trace transformation of Section~\ref{sec:bc-schur} to the two
line-defect integrals of \cite[Sec.~5.4.1]{Hatsuda:2025jze},
with a pair of fundamental $USp(2n)$ Wilson lines on one side and two
minuscule 't~Hooft lines of $SO(2n+1)$, at the appropriate discrete
theta angle, on the other, and similarly to the Wilson and vortex
lines of Hatsuda and Okazaki \cite{Hatsuda:2026xxp}.

In three dimensions, mirror symmetry
\cite{Intriligator:1996ex,Hanany:1996ie} gives many further
identities: the index identity
for the basic abelian pair is \cite[Eq.~(4.20)]{Hwang:2012jh},
mirror pairs obtained by gauging flavor symmetries of linear quivers
were constructed in \cite{Dey:2014tka}, and indices and
half-indices of many pairs were compared in
\cite{Okazaki:2019ony,Okazaki:2026epb}. The sequences of local
dualities of Hwang, Pasquetti and Sacchi \cite{Hwang:2021ulb}
suggest a route to all-orders identities for nonabelian mirror pairs,
including the discretely gauged $T[SU(3)]$ theories of
\cite[Eq.~(181)]{Nawata:2023rdx}. For the $(A_1,D_5)$ example
of this paper, the circle reductions of both descriptions are expected
to give $T[SU(2)]$ together with a free twisted hypermultiplet
\cite[Sec.~4]{Cho:2026tqf}, which could be tested at the level of
three-dimensional indices.

\vspace{0.2in} {\leftline {\bf Acknowledgments}}
\nopagebreak
We thank the participants of the October 2026 workshop
\emph{Supersymmetric Quantum Field Theories Across Dimensions} at the
Peng Huanwu Center for Fundamental Theory, University of Science and
Technology of China (USTC), for helpful discussions. We developed the
derivations and proofs with assistance from AI tools including GPT-6 Astra, GPT-6.2 Astra.
We also used AI tools in writing the manuscript. We take responsibility for the
arguments and the final text. This work was supported by the National
Natural Science Foundation of China under Grants No.~12325502 and
No.~12247103. XW is also supported by the Fundamental Research Funds for the Central Universities 
(Grants No. WK2030250140)

\let\originalthebibliography\thebibliography
\renewcommand{\thebibliography}[1]{%
  \originalthebibliography{#1}%
  \addcontentsline{toc}{section}{\refname}%
  \setlength{\itemsep}{3pt plus 1pt minus 1pt}%
}
\bibliographystyle{JHEP}
\bibliography{referenceSCI}
\end{document}